\documentclass[preprint,prd,aps,showpacs,showkeys,nofootinbib]{revtex4}
\usepackage{graphicx}
\usepackage{dcolumn}
\usepackage{bm}
\begin{document}

%\begin{flushright}
%{hep-ph/0405xxx}\\
%\end{flushright}
%\preprint{hep-ph/0405192}

\title{Heavy fermions and two loop electroweak corrections to $b\rightarrow s+\gamma$}

\author{Xiu-Yi Yang$^1$, Tai-Fu Feng$^{1,2}$}

\affiliation{$^1$Department of Physics, Dalian University of
Technology, Dalian 116024, China\\
$^2$Center for High Energy Physics, Peking University,
Beijing 100871, China}
\date{\today}

\begin{abstract}
Applying effective Lagrangian method and on-shell scheme,
we analyze the electroweak corrections to the rare decay $b\rightarrow s+\gamma$
from some special two loop diagrams in which a closed
heavy fermion loop is attached to the virtual charged gauge bosons or Higgs.
At the decoupling limit where the virtual fermions in
inner loop are much heavier than the electroweak scale, we verify
the final results satisfying the decoupling theorem explicitly
when the interactions among Higgs and heavy fermions do not contain
the nondecoupling couplings. Adopting the universal assumptions on
the relevant couplings and mass spectrum of new physics, we find that
the relative corrections from those two loop diagrams to the
SM theoretical prediction on the branching ratio of $B\rightarrow X_{_s}\gamma$
can reach $5\%$ as the energy scale of new physics $\Lambda_{_{\rm NP}}=200$ GeV.
\end{abstract}



\maketitle

\section{Introduction}
\indent\indent
The rare $B$ decays serve as a good test for new physics beyond the standard model
(SM) since they are not seriously affected by the uncertainties originating from
long distance effects. The forthcoming and running $B$ factories will make more
precise measurements on the rare $B$-decay processes, and those measurements should
set more strict constraints on the new physics beyond SM. The main purpose of
investigating $B$-decay, especially the rare decay modes, is to search for
traces of new physics and determines its parameter space.

The measurements of the branching ratios at CLEO, ALEPH and BELLE
\cite{exp} give the combined result
\begin{equation}
BR(B\rightarrow X_{_s}\gamma)=(3.52\pm0.23\pm0.09)\times10^{-4}\;,
\label{eq1}
\end{equation}
which agrees with the next-to-next-to-leading order (NNLO) standard model
(SM) prediction \cite{smp}
\begin{equation}
BR(B\rightarrow X_{_s}\gamma)=(3.15\pm0.23)\times10^{-4}\;.
\label{eq2}
\end{equation}
Good agreement between the experiment and the theoretical prediction
of the SM implies that the new physics scale should lie well above
the electroweak (EW) scale $\Lambda_{_{\rm EW}}$. The systematic analysis of
new physics corrections to $B\rightarrow X_{_s}\gamma$
up to two-loop order can help us understanding where the
new physics scale sets in, and the distribution of new physical
particle masses around this scale. In principle, the two-loop
corrections can be large when some additional parameters are involved
at this perturbation order besides the parameters appearing in
one loop results. In other words, including the two-loop
contributions one can obtain a more exact constraint on the new physics
parameter space from the present experimental results.

Though the QCD corrections to the rare B decays are discussed extensively
in literature, the pure two-loop EW corrections to the branching
ratio of $b\rightarrow s\gamma$ are less advanced because of the well known
difficulty in calculation. Strumia has evaluated the two-loop EW corrections
to $b\rightarrow s\gamma$ from the top quark using heavy mass expansion
in gaugeless limit of the SM \cite{Strumia}.
At the limit of large $\tan\beta$ in supersymmetry, Ref.\cite{Borzumati}
analyzes the two loop corrections to the branching ratio of
$B\rightarrow X_{_s}\gamma$ from the virtual charged Higgs
and gluino-squark sector.

Employing the effective Lagrangian method and on-shell scheme, we present
the corrections to the branch ratio of $B\rightarrow X_{_s}\gamma$ from
some special diagrams in which a closed heavy
fermion loop is attached to the virtual charged gauge bosons or Higgs here.
The effective Lagrangian method can yield one loop EW corrections to
the effective Lagrangian of $b\rightarrow s\gamma$ exactly
in the SM and beyond, and has been adopted to calculate the two loop supersymmetric corrections
for the branching ratio of $b\rightarrow s\gamma$ \cite{Feng1}, neutron EDM \cite{Feng2}
and lepton MDMs and EDMs \cite{Feng3,Feng4}. In concrete
calculation, we assume that all external quarks and photon
are off-shell, then expand the amplitude of corresponding triangle
diagrams according to the external momenta of quarks and photon.
Using loop momentum translating invariance, we formulate the sum of
amplitude from those triangle diagrams corresponding to same self energy in
the form which explicitly satisfies the Ward identity required by
the QED gauge symmetry, then get all dimension 6 operators
together with their coefficients. After the equations of
motion are applied to external quarks, higher dimensional operators, such
as dimension 8 operators, also contribute to the branching ratio of
$B\rightarrow X_{_s}\gamma$ in principle. However, the contributions of dimension 8
operators contain the additional suppression factor
$m_{_b}^2/\Lambda_{_{\rm EW}}^2$ comparing with that of dimension
6 operators, where $m_{_b}$ is the mass of bottom quark. Setting
$\Lambda_{_{\rm EW}}\sim100{\rm GeV}$, one obtains easily that this
suppression factor is about $10^{-3}$ for the $b\rightarrow s\gamma$.
Under current experimental precision, it implies
that the contributions of all higher dimension operators ($D\ge8$)
can be neglected safely.

We adopt the naive dimensional regularization with the
anticommuting $\gamma_{_5}$ scheme, where there is no distinction
between the first 4 dimensions and the remaining $D-4$ dimensions.
Since the bare effective Lagrangian contains the ultraviolet
divergence which is induced by divergent subdiagrams, we give the
renormalized results in the on-mass-shell scheme \cite{onshell}.
Additional, we adopt the nonlinear $R_\xi$ gauge with $\xi=1$ for
simplification \cite{nonlinear-R-xi}. This special gauge-fixing term
guarantees explicit electromagnetic gauge invariance throughout the calculation,
not just at the end because the choice of gauge-fixing term
eliminates the $\gamma W^\pm G^\mp$ vertex in the Lagrangian.

This paper is composed of the sections as follows.
In section \ref{sec2}, we introduce the effective Lagrangian
method and our notations. We will demonstrate how to obtain
the identities among two loop integrals from the loop momentum
translating invariance through an example, then obtain the
corrections from the relevant diagrams to the effective Lagrangian
of $b\rightarrow s\gamma$. Section \ref{sec3} is devoted to the numerical
discussion under universal assumptions on the parameters of new physics.
In section \ref{sec4}, we give our conclusion. Some
tedious formulae are collected in the appendices.

\section{The Wilson coefficients from the two-loop diagrams\label{sec2}}
\indent\indent

In this section, we derive the relevant Wilson coefficients
for the partonic decay $b\rightarrow s\gamma$ including two-loop
EW corrections. In a conventional form, the effective Hamilton is
written as
\begin{eqnarray}
&&H_{_{eff}}= -{4G_{_F}\over\sqrt{2}}V_{_{ts}}^*V_{_{tb}}
\sum\limits_{i}C_{_i}(\mu){\cal O}_{_i}\;,
\label{eq3}
\end{eqnarray}
where $V$ is the CKM matrix and $G_{_F}=1.16639\times10^{-5}\;{\rm GeV}^{-2}$
is the 4-fermion coupling. The definitions of those dimension six
operators are \cite{Buras1}
\begin{eqnarray}
&&{\cal O}_{_1}={1\over(4\pi)^2}\;\bar{s}(i/\!\!\!\!{\cal D})^3
\omega_-b\;,\nonumber\\
&&{\cal O}_{_2}={eQ_{_d}\over(4\pi)^2}\Big[\overline{(i{\cal D}_{_\mu}s)}
\gamma^\mu F\cdot\sigma\omega_- b+\bar{s}F\cdot\sigma\gamma^\mu
\omega_-(i{\cal D}_{_\mu}b)\Big]\;,\nonumber\\
&&{\cal O}_{_3}={eQ_{_d}\over(4\pi)^2}\;\bar{s}(\partial^\mu F_{_{\mu\nu}})
\gamma^\nu\omega_- b\;,\nonumber\\
&&{\cal O}_{_4}={1\over(4\pi)^2}\;\bar{s}(i/\!\!\!\!{\cal D})^2
\Big(m_{_b}\omega_++m_{_s}\omega_-\Big)b\;,\nonumber\\
&&{\cal O}_{_5}={eQ_{_d}\over(4\pi)^2}\;\bar{s}\sigma^{\mu\nu}
\Big(m_{_b}\omega_++m_{_s}\omega_-\Big)bF_{_{\mu\nu}}\;,\nonumber\\
&&{\cal O}_{_6}={g_{_s}\over(4\pi)^2}\Big[\overline{(i{\cal D}_{_\mu}s)}
\gamma^\mu G\cdot\sigma\omega_- b+\bar{s}G\cdot\sigma\gamma^\mu
\omega_-(i{\cal D}_{_\mu}b)\Big]\;,\nonumber\\
&&{\cal O}_{_{7}}={g_{_s}\over(4\pi)^2}\;\bar{s}(\partial^\mu G_{_{\mu\nu}})
\gamma^\nu\omega_- b\;,\nonumber\\
&&{\cal O}_{_{8}}={g_{_s}\over(4\pi)^2}\;\bar{s}T^a\sigma^{\mu\nu}
\Big(m_{_b}\omega_++m_{_s}\omega_-\Big)bG^a_{_{\mu\nu}}\;,\nonumber\\
&&{\cal O}_{_9}=-{eQ_{_d}\over(4\pi)^2}\Big[\overline{(i{\cal D}_{_\mu}s)}
\gamma^\mu F\cdot\sigma\omega_- b-\bar{s}F\cdot\sigma\gamma^\mu
\omega_-(i{\cal D}_{_\mu}b)\Big]\;,\nonumber\\
&&{\cal O}_{_{10}}={1\over(4\pi)^2}\;\bar{s}(i/\!\!\!\!{\cal D})^2
\Big(m_{_b}\omega_+-m_{_s}\omega_-\Big)b\;,\nonumber\\
&&{\cal O}_{_{11}}={eQ_{_d}\over(4\pi)^2}\;\bar{s}\sigma^{\mu\nu}
\Big(m_{_b}\omega_+-m_{_s}\omega_-\Big)bF_{_{\mu\nu}}\;,\nonumber\\
&&{\cal O}_{_{12}}=-{g_{_s}\over(4\pi)^2}\Big[\overline{(i{\cal D}_{_\mu}s)}
\gamma^\mu G\cdot\sigma\omega_- b-\bar{s}G\cdot\sigma\gamma^\mu
\omega_-(i{\cal D}_{_\mu}b)\Big]\;,\nonumber\\
&&{\cal O}_{_{13}}={g_{_s}\over(4\pi)^2}\;\bar{s}T^a\sigma^{\mu\nu}
\Big(m_{_b}\omega_+-m_{_s}\omega_-\Big)bG^a_{_{\mu\nu}}\;,\nonumber\\
&&{\cal O}_{_{14}}=(\bar{s}_{_\alpha}\gamma^\mu\omega_-c_{_\alpha})
(\bar{c}_{_\beta}\gamma^\mu\omega_-b_{_\beta})\;,
\label{eq4}
\end{eqnarray}
where $F_{_{\mu\nu}}$ and $G_{_{\mu\nu}}=G^a_{_{\mu\nu}}T^a$ are the field strengths of the
photon and gluon respectively, and $T^a\;(a=1,\;\cdots,\;8)$ are $SU(3)_{_c}$
generators. In addition, $e$ and $g_{_s}$ represent the EW and strong couplings
respectively.

After expanding the amplitude of corresponding triangle diagrams,
we extract the Wilson coefficients of operators in Eq.(\ref{eq4})
which are formulated in the linear combinations of one and two loop
vacuum integrals in momentum space, then obtain the corrections to
the branching ratio of $B\rightarrow X_{_s}\gamma$.
Taking those diagrams in which a closed heavy fermion loop is inserted
into the propagator of charged gauge boson as an example, we show in detail
how to obtain the Wilson coefficients in effective Lagrangian.
%%%%%%%%%%%%%%%%%%%%%%%%%%%END MODIFICATION%%%%%%%%%%%%%%%%%%%%%%%%%%%%%

\subsection{The corrections from the diagrams where a closed heavy fermion
loop is inserted into the self energy of $W^\pm$ gauge boson}
\begin{figure}[t]
\setlength{\unitlength}{1mm}
\begin{center}
\begin{picture}(0,40)(0,0)
\put(-60,-110){\includegraphics{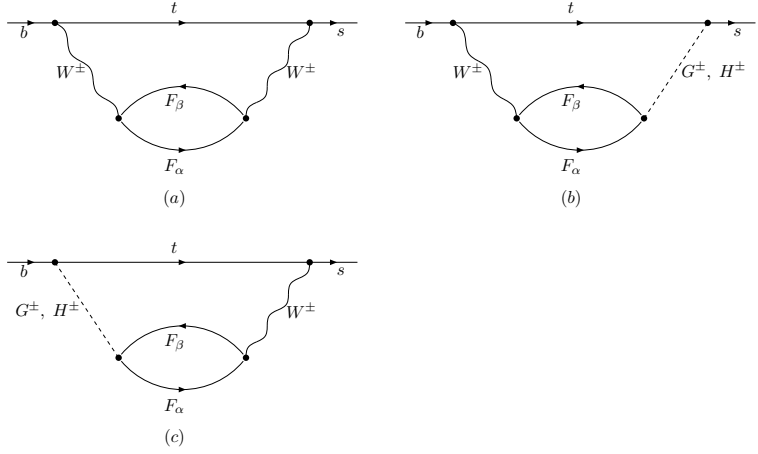}}
\end{picture}
\caption[]{The relating two-loop diagrams in which a closed heavy
fermion loop is attached to virtual $W^\pm$ bosons or $G^\pm\;(H^\pm)$,
where a real photon or gluon is attached in all possible way.}
\label{fig1}
\end{center}
\end{figure}

\indent\indent
In order to get the amplitude of the diagrams in Fig.\ref{fig1}(a), one can
write the renormalizable interaction among the charged EW gauge
boson $W^\pm$ and the heavy fermions $F_{\alpha,\beta}$ in a more universal
form as
\begin{eqnarray}
&&{\cal L}_{_{WFF}}={e\over s_{_{\rm w}}}W^{-,\mu}\bar{F}_\alpha\gamma_\mu
(\zeta^L_{_{\alpha\beta}}\omega_-+\zeta^R_{_{\alpha\beta}}\omega_+)F_\beta
+h.c.\;,
\label{WFF}
\end{eqnarray}
where the concrete expressions of $\zeta^{L,R}_{_{\alpha\beta}}$ depend on the
models employed in our calculation. The conservation of electric charge
requires $Q_\beta-Q_\alpha=1$, where $Q_{\alpha,\beta}$ denote the electric
charge of the heavy fermions $F_{\alpha,\beta}$ respectively.

%%%%%%%%%%%%%%%%%%%%%%%%%%%%%%%%%%%%%%%%%%%%%%%%%%%%%%%%%%%%%%%%%%%
%%%%%%%%%%%%%%%%%%%%%%%%%%%%%%%%%%%%%%%%%%%%%%%%%%%%%%%%%%%%%%%%%%%

%%%%%%%%%%%%%%%%%%%%%%%%%%%%%%%%%%%%%%%%%%%%%%%%%%%%%%%%%%%%%%%%%%%
\begin{figure}[t]
\setlength{\unitlength}{1mm}
\begin{center}
\begin{picture}(0,40)(0,0)
\put(-40,-130){\includegraphics{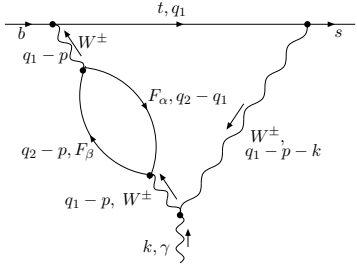}}
\end{picture}
\caption[]{The triangle diagram in which the real photon is attached
to $W^\pm$ gauge boson. The amplitude is written in Eq(\ref{eq-wa1}).}
\label{fig2}
\end{center}
\end{figure}
%%%%%%%%%%%%%%%%%%%%%%%%%%%%%%%%%%%%%%%%%%%%%%%%%%%%%%%%%%%%%%%%%%%
%%%%%%%%%%%%%%%%%%%%%%%%%BEGIN MODIFICATION%%%%%%%%%%%%%%%%%%%%%%%%%%%%%
Applying Eq.(\ref{WFF}), we write firstly the amplitude of those two loop
diagrams in Fig.\ref{fig1}(a). For example, the amplitude for the diagram
in which a real photon is attached to the virtual $W^\pm$ boson (Fig.\ref{fig2})
can be formulated as
\begin{eqnarray}
%%%%%%%%%%%%%%%%%%%%%%%%%%%%%%%%%%%%%%%%%%%%%%%%%%%%%%%%%%%%%%%%%%%%%%%%%%%%%%%%%%%
&&i{\cal A}_{_{{\rm ww},\rho}}^{(\ref{fig2})}(p,k)=-\overline{\psi}_{_s}\int{d^D q_1\over(2\pi)^D}
{d^D q_2\over(2\pi)^D}\Big(-i{e\Lambda_{_{\rm RE}}^\varepsilon\over\sqrt{2}
s_{_{\rm w}}}V_{_{ts}}^*\Big)\gamma^\mu\omega_-{i/\!\!\!q_1+m_{_t}\over
q_1^2-m_{_t}^2}\Big(-i{e\Lambda_{_{\rm RE}}^\varepsilon\over
\sqrt{2}s_{_{\rm w}}}V_{_{tb}}\Big)\gamma^\nu\omega_-\psi_{_b}
\nonumber\\
&&\hspace{2.8cm}\times
{-i\over(q_1-p-k)^2-m_{_{\rm w}}^2}
\Big\{ie\Big[-g_{\mu\sigma}(2p+k-2q_1)_\rho
+2(g_{\rho\mu}k_\sigma-g_{\rho\sigma}k_\mu)\Big]\Big\}
\nonumber\\
&&\hspace{2.8cm}\times
{-i\over(q_1-p)^2-m_{_{\rm w}}^2}{-i\over(q_1-p)^2-m_{_{\rm w}}^2}
{\bf Tr}\Bigg[\Big(i{e\Lambda_{_{\rm RE}}^\varepsilon\over s_{_{\rm w}}}\Big)
\gamma^\sigma\Big\{\zeta^{L*}_{_{\alpha\beta}}\omega_-
+\zeta^{R*}_{_{\alpha\beta}}\omega_+\Big\}
\nonumber\\
&&\hspace{2.8cm}\times
{i(/\!\!\!q_2-/\!\!\!q_1+m_{_{F_\alpha}})
\over(q_2-q_1)^2-m_{_{F_\alpha}}^2}\Big(i{e\Lambda_{_{\rm RE}}^\varepsilon\over
s_{_{\rm w}}}\Big)\gamma_\nu\Big\{\zeta^L_{_{\alpha\beta}}\omega_-
+\zeta^R_{_{\alpha\beta}}\omega_+\Big\}
{i(/\!\!\!q_2-/\!\!\!\!p+m_{_{F_\beta}})\over(q_2-p)^2-m_{_{F_\beta}}^2}\Bigg]\;.
%%%%%%%%%%%%%%%%%%%%%%%%%%%%%%%%%%%%%%%%%%%%%%%%%%%%%%%%%%%%%%%%%%%%%%%%%%%%%%%%%%%
\label{eq-wa1}
\end{eqnarray}
Here $\Lambda_{_{\rm RE}}$ denotes the renormalization scale that can take
any value in the range from the EW scale $\Lambda_{_{\rm EW}}$ to the
new physics scale $\Lambda_{_{\rm NP}}$ naturally, and
we adopt the abbreviations: $c_{_{\rm w}}=\cos\theta_{_{\rm w}},\;s_{_{\rm w}}
=\sin\theta_{_{\rm w}}$ with $\theta_{_{\rm w}}$ denoting the Weinberg angle.
Additionally, $p,\;k$ are the incoming momenta of quark and photon fields,
$\rho$ is the Lorentz index of photon, separatively. Certainly,
the amplitude does not depend on how to mark the momenta of virtual fields because of
the translating invariance of loop momenta.

It can be checked easily that the sum of amplitude for diagrams in Fig.\ref{fig1}(a)
satisfies the Ward identity required by the QED gauge invariance
\begin{eqnarray}
&&k^\rho{\cal A}_{_{{\rm ww},\rho}}^{(\ref{fig1}(a))}(p,k)
=e[\Sigma_{_{\rm ww}}^{(\ref{fig1}(a))}(p+k)
-\Sigma_{_{\rm ww}}^{(\ref{fig1}(a))}(p)]\;,
\label{WTI-ww}
\end{eqnarray}
where ${\cal A}_{_{{\rm ww},\rho}}^{(\ref{fig1}(a))}$
denotes the sum of amplitudes for the triangle diagrams corresponding to
the self energy in Fig.\ref{fig1}(a), as well as $\Sigma_{_{\rm ww}}^{(\ref{fig1}(a))}$
denotes the amplitude of corresponding self energy
diagram, respectively.

According the external momenta of quarks and photon, we expand
the amplitude in Eq.(\ref{eq-wa1}) as
\begin{eqnarray}
%%%%%%%%%%%%%%%%%%%%%%%%%%%%%%%%%%%%%%%%%%%%%%%%%%%%%%%%%%%%%%%%%%%%%%%%%%%%%%%%%%%
&&i{\cal A}_{_{{\rm ww},\rho}}^{(\ref{fig2})}(p,k)
=-i{e^5\over2s_{_{\rm w}}^4}V_{_{ts}}^*V_{_{tb}}\cdot\Lambda_{_{\rm RE}}^{4\epsilon}
\int{d^D q_1\over(2\pi)^D}{d^Dq_2\over(2\pi)^D}{1\over {\cal D}_{_{\rm ww}}}
\Bigg\{1+{2q_1\cdot(3p+k)\over q_1^2-m_{_{\rm w}}^2}
\nonumber\\
&&\hspace{2.8cm}
+{2q_1\cdot p\over q_2^2-m_{_{F_\beta}}^2}-{2p^2+(p+k)^2\over q_1^2-m_{_{\rm w}}^2}
-{p^2\over q_2^2-m_{_{F_\beta}}^2}+{4(q_2\cdot p)^2\over(q_2^2-m_{_{F_\beta}}^2)^2}
\nonumber\\
&&\hspace{2.8cm}
+{4(q_1\cdot(p+k))^2+8(q_1\cdot p)(q_1\cdot(p+k))+12(q_1\cdot p)^2
\over(q_1^2-m_{_{\rm w}}^2)^2}
\nonumber\\
&&\hspace{2.8cm}
+{4(q_1\cdot(3p+k))(q_2\cdot p)\over(q_1^2-m_{_{\rm w}}^2)
(q_2^2-m_{_{F_\beta}}^2)}\Bigg\}
\overline{\psi}_{_s}\Big[\gamma^\mu/\!\!\!q_1\gamma^\nu\omega_-\Big]\psi_{_b}
\Big[-g_{\mu\sigma}(2p+k-2q_1)_\rho
\nonumber\\
&&\hspace{2.8cm}
+2(g_{\rho\mu}k_\sigma-g_{\rho\sigma}k_\mu)\Big]
{\bf Tr}\Bigg[\gamma^\sigma\Big\{\zeta^{L*}_{_{\alpha\beta}}\omega_-
+\zeta^{R*}_{_{\alpha\beta}}\omega_+\Big\}
(/\!\!\!q_2-/\!\!\!q_1+m_{_{F_\alpha}})
\nonumber\\
&&\hspace{2.8cm}\times
\gamma_\nu\Big\{\zeta^L_{_{\alpha\beta}}\omega_-+\zeta^R_{_{\alpha\beta}}\omega_+\Big\}
(/\!\!\!q_2-/\!\!\!\!p+m_{_{F_\beta}})\Bigg]
%%%%%%%%%%%%%%%%%%%%%%%%%%%%%%%%%%%%%%%%%%%%%%%%%%%%%%%%%%%%%%%%%%%%%%%%%%%%%%%%%%%
\label{eq-wa2}
\end{eqnarray}
since we only consider the corrections from dimension 6 operators,
here ${\cal D}_{_{\rm ww}}=(q_1^2-m_{_t}^2)
(q_1^2-m_{_{\rm w}}^2)^3((q_2-q_1)^2-m_{_{F_\alpha}}^2)(q_2^2-m_{_{F_\beta}}^2)$.

Because the denominators of all terms in Eq.(\ref{eq-wa2}) are invariant under the reversal
$q_1\rightarrow-q_1,q_2\rightarrow-q_2$, those terms in odd powers of
loop momenta can be abandoned, and the terms in even powers of loop
momenta can be simplified by
\begin{eqnarray}
%%%%%%%%%%%%%%%%%%%%%%%%%%%%%%%%%%%%%%%%%%%%%%%%%%%%%%%%%%%%%%%%%%%%%%%%%%%%%%%%%%%
&&\int{d^Dq_1\over(2\pi)^D}{d^Dq_2\over(2\pi)^D}{q_{1\mu}q_{1\nu}q_{1\rho}
q_{1\sigma}q_{1\alpha}q_{1\beta},\;q_{1\mu}q_{1\nu}q_{1\rho}
q_{1\sigma}q_{1\alpha}q_{2\beta}\over((q_2-q_1)^2-m_{_0}^2)(q_1^2-m_{_1}^2)(q_2^2-m_{_2}^2)}
\nonumber\\
&&\hspace{-1.0cm}
\longrightarrow{S_{_{\mu\nu\rho\sigma\alpha\beta}}\over D(D+2)(D+4)}
\int{d^Dq_1\over(2\pi)^D}{d^Dq_2\over(2\pi)^D}{(q_1)^3,\;(q_1)^2q_1\cdot q_2
\over((q_2-q_1)^2-m_{_0}^2)(q_1^2-m_{_1}^2)(q_2^2-m_{_2}^2)}
\;,\nonumber\\
%%%%%%%%%%%%%%%%%%%%%%%%%%%%%%%%%%%%%%%%%%%%%%%%%%%%%%%%%%%%%%%%%%%%%%%%%%%%%%%%%%%
&&\int{d^Dq_1\over(2\pi)^D}{d^Dq_2\over(2\pi)^D}{q_{1\mu}q_{1\nu}q_{1\rho}
q_{1\sigma}q_{2\alpha}q_{2\beta}\over((q_2-q_1)^2-m_{_0}^2)(q_1^2-m_{_1}^2)(q_2^2-m_{_2}^2)}
\nonumber\\
&&\hspace{-1.0cm}
\longrightarrow\int{d^Dq_1\over(2\pi)^D}{d^Dq_2\over(2\pi)^D}
{1\over((q_2-q_1)^2-m_{_0}^2)(q_1^2-m_{_1}^2)
(q_2^2-m_{_2}^2)}
\nonumber\\
&&\hspace{-0.2cm}\times
\Big[{Dq_{_1}^2(q_{_1}\cdot q_{_2})^2-(q_{_1}^2)^2q_{_2}^2\over D(D-1)(D+2)(D+4)}
S_{_{\mu\nu\rho\sigma\alpha\beta}}
-{q_{_1}^2(q_{_1}\cdot q_{_2})^2-(q_{_1}^2)^2q_{_2}^2
\over D(D-1)(D+2)}T_{_{\mu\nu\rho\sigma}}g_{_{\alpha\beta}}\Big]
\;,\nonumber\\
%%%%%%%%%%%%%%%%%%%%%%%%%%%%%%%%%%%%%%%%%%%%%%%%%%%%%%%%%%%%%%%%%%%%%%%%%%%%%%%%%%%
&&\int{d^Dq_1\over(2\pi)^D}{d^Dq_2\over(2\pi)^D}{q_{1\mu}q_{1\nu}q_{1\rho}
q_{2\alpha}q_{2\beta}q_{2\delta}\over((q_2-q_1)^2-m_{_0}^2)(q_1^2-m_{_1}^2)(q_2^2-m_{_2}^2)}
\nonumber\\
&&\hspace{-1.0cm}
\longrightarrow\int{d^Dq_1\over(2\pi)^D}{d^Dq_2\over(2\pi)^D}
{1\over((q_2-q_1)^2-m_{_0}^2)(q_1^2-m_{_1}^2)
(q_2^2-m_{_2}^2)}
\nonumber\\
&&\hspace{-0.2cm}\times
\Big[{(D+1)q_{_1}^2q_{_1}\cdot q_{_2}q_{_2}^2-2(q_{_1}\cdot q_{_2})^3
\over D(D-1)(D+2)(D+4)}S_{_{\mu\nu\rho\alpha\beta\delta}}
+{(q_{_1}\cdot q_{_2})^3-q_{_1}^2q_{_1}\cdot q_{_2}q_{_2}^2\over
D(D-1)(D+2)}\Big(g_{_{\mu\alpha}}(g_{_{\nu\beta}}g_{_{\rho\delta}}
\nonumber\\
&&\hspace{-0.2cm}
+g_{_{\nu\delta}}g_{_{\rho\beta}})+g_{_{\mu\beta}}(g_{_{\nu\alpha}}
g_{_{\rho\delta}}+g_{_{\nu\delta}}g_{_{\rho\alpha}})
+g_{_{\mu\delta}}(g_{_{\nu\alpha}}g_{_{\rho\beta}}
+g_{_{\nu\beta}}g_{_{\rho\alpha}})\Big)\Big]\;,
%%%%%%%%%%%%%%%%%%%%%%%%%%%%%%%%%%%%%%%%%%%%%%%%%%%%%%%%%%%%%%%%%%%%%%%%%%%%%%%%%%%
\label{eq-wa3}
\end{eqnarray}
and those similar formulae presented in Eq.(5) of Ref\cite{Feng1},
where the tensors are defined as
\begin{eqnarray}
%%%%%%%%%%%%%%%%%%%%%%%%%%%%%%%%%%%%%%%%%%%%%%%%%%%%%%%%%%%%%%%%%%%%%%%%%%%%%%%%%%%
&&T_{_{\mu\nu\rho\sigma}}=g_{_{\mu\nu}}g_{_{\rho\sigma}}+g_{_{\mu\rho}}g_{_{\nu\sigma}}
+g_{_{\mu\sigma}}g_{_{\nu\rho}}
\;,\nonumber\\
%%%%%%%%%%%%%%%%%%%%%%%%%%%%%%%%%%%%%%%%%%%%%%%%%%%%%%%%%%%%%%%%%%%%%%%%%%%%%%%%%%%
&&S_{_{\mu\nu\rho\sigma\alpha\beta}}=
g_{_{\mu\nu}}T_{_{\rho\sigma\alpha\beta}}+g_{_{\mu\rho}}T_{_{\nu\sigma\alpha\beta}}
+g_{_{\mu\sigma}}T_{_{\nu\rho\alpha\beta}}+g_{_{\mu\alpha}}T_{_{\nu\rho\sigma\beta}}
+g_{_{\mu\beta}}T_{_{\nu\rho\sigma\alpha}}\;.
%%%%%%%%%%%%%%%%%%%%%%%%%%%%%%%%%%%%%%%%%%%%%%%%%%%%%%%%%%%%%%%%%%%%%%%%%%%%%%%%%%%
\label{eq-wa4}
\end{eqnarray}
Summing over those indices which appear both as superscripts and subscripts simultaneously,
we derive all possible dimension 6 operators in the momentum space together with their
coefficients which are expressed in the linear combinations of one and
two loop vacuum integrals. In a similar way, one obtains the amplitude
of other diagrams. Before integrating with the loop momenta, we apply the loop momentum
translating invariance to formulate the sum of those amplitude in explicitly
QED gauge invariant form, then extract the Wilson coefficients of those
dimension 6 operators listed in Eq.(\ref{eq4}). Actually, we can easily
verify the equation
\begin{eqnarray}
&&\int\int{d^Dq_1\over(2\pi)^D}{d^Dq_2\over(2\pi)^D}{q_{1\mu}\over
(q_1^2-m_1^2)(q_2^2-m_2^2)((q_2-q_1)^2-m_0^2)}\equiv0\;.
\label{l-tran1}
\end{eqnarray}
Performing an infinitesimal translation $q_1\rightarrow q_1,\;q_2\rightarrow
q_2-a$ with $a_\rho\rightarrow0\;(\rho=0,1,\cdots,D)$, one can write the
left-handed side of above equation as
\begin{eqnarray}
&&\int\int{d^Dq_1\over(2\pi)^D}{d^Dq_2\over(2\pi)^D}{q_{1\mu}\over
(q_1^2-m_1^2)(q_2^2-m_2^2)((q_2-q_1)^2-m_0^2)}
\nonumber\\
&&\hspace{-0.6cm}=
\int\int{d^Dq_1\over(2\pi)^D}{d^Dq_2\over(2\pi)^D}{q_{1\mu}\over
(q_1^2-m_1^2)(q_2^2-m_2^2)((q_2-q_1)^2-m_0^2)}
\nonumber\\
&&\hspace{-0.2cm}\times
\Big\{1+{2q_2\cdot a\over q_2^2-m_2^2}+{2(q_2-q_1)\cdot a\over(q_2-q_1)^2-m_0^2}+\cdots\Big\}\;.
\label{l-tran2}
\end{eqnarray}
This result implies
\begin{eqnarray}
&&\int\int{d^Dq_1\over(2\pi)^D}{d^Dq_2\over(2\pi)^D}{q_1\cdot q_2\over
(q_1^2-m_1^2)(q_2^2-m_2^2)^2((q_2-q_1)^2-m_0^2)}
\nonumber\\
&&\hspace{-0.6cm}=
\int\int{d^Dq_1\over(2\pi)^D}{d^Dq_2\over(2\pi)^D}{q_1^2-q_1\cdot q_2\over
(q_1^2-m_1^2)(q_2^2-m_2^2)((q_2-q_1)^2-m_0^2)^2}\;.
\label{l-tran3}
\end{eqnarray}
In a similar way, other identities presented in Ref.\cite{Feng1} can be derived.
Using the expression of two loop vacuum integral\cite{Davydychev}
\begin{eqnarray}
&&\Lambda_{_{\rm RE}}^{4\epsilon}\int\int{d^Dq_1\over(2\pi)^D}{d^Dq_2\over(2\pi)^D}{1\over
(q_1^2-m_1^2)(q_2^2-m_2^2)((q_2-q_1)^2-m_0^2)}
\nonumber\\
&&\hspace{-0.6cm}=
{\Lambda^2\over2(4\pi)^4}{\Gamma^2(1+\epsilon)\over(1-\epsilon)^2}
\Big({4\pi x_{_R}}\Big)^{2\epsilon}
\Big\{-{1\over\epsilon^2}\Big(x_0+x_1+x_2\Big)
\nonumber\\&&
+{1\over\epsilon}\Big(2(x_0\ln x_0+x_1\ln x_1+x_2\ln x_2)-x_0-x_1-x_2\Big)
\nonumber\\&&
-2(x_0+x_1+x_2)+2(x_0\ln x_0+x_1\ln x_1+x_2\ln x_2)
\nonumber\\&&
-x_0\ln^2x_0-x_1\ln^2x_1-x_2\ln^2x_2-\Phi(x_0,x_1,x_2)\Big\}
\label{2l-vacuum}
\end{eqnarray}
and
\begin{eqnarray}
&&\Phi(x,y,z)=(x+y-z)\ln x\ln y+(x-y+z)\ln x\ln z
\nonumber\\
&&\hspace{2.2cm}
+(-x+y+z)\ln y\ln z+{\rm sign}(\lambda^2)\sqrt{|\lambda^2|}\Psi(x,y,z)\;,
\nonumber\\
&&{\partial\Phi\over\partial x}(x,y,z)=\ln x\ln y+\ln x\ln z
-\ln y\ln z+2\ln x+{x-y-z\over\sqrt{|\lambda^2|}}\Psi(x,y,z)\;,
\label{phi}
\end{eqnarray}
one obtains easily
\begin{eqnarray}
&&{\Lambda_{_{\rm RE}}^{4\epsilon}\over\Lambda^2}{\partial\over\partial x_0}
\bigg\{\int\int{d^Dq_1\over(2\pi)^D}{d^Dq_2\over(2\pi)^D}
{q_1^2\over(q_1^2-m_1^2)(q_2^2-m_2^2)((q_2-q_1)^2-m_0^2)}\bigg\}
\nonumber\\
%%%%%%%%%%%%%%%%%%%%%%%%%%%%%%%%%%%%%%%%%%%%%%%%%%%%%%%%%%%%%%%%%%%%%%%%%%%%
&&\hspace{-0.5cm}=
{\Lambda_{_{\rm RE}}^{4\epsilon}\over\Lambda^2}
\Big\{{\partial\over\partial x_0}+{\partial\over\partial x_2}\Big\}
\bigg\{\int\int{d^Dq_1\over(2\pi)^D}{d^Dq_2\over(2\pi)^D}{q_1\cdot q_2\over
(q_1^2-m_1^2)(q_2^2-m_2^2)((q_2-q_1)^2-m_0^2)}\bigg\}
\nonumber\\
&&\hspace{-0.5cm}=
{\Lambda^2\over2(4\pi)^4}{\Gamma^2(1+\epsilon)\over(1-\epsilon)^2}
\Big({4\pi x_{_R}}\Big)^{2\epsilon}\Big\{-{x_1+2x_2\over\epsilon^2}
+{1\over\epsilon}\Big(x_1(1+2\ln x_0)+2x_2(1+\ln x_0+\ln x_2)\Big)
\nonumber\\&&
-(x_1+x_2)\ln^2x_0-(x_1+2x_2)\ln x_0\ln x_2-x_2\ln^2x_2-x_1\ln x_0\ln x_1
+x_1\ln x_1\ln x_2
\nonumber\\&&
-2(x_1+x_2)\ln x_0-2x_2\ln x_2-{x_1(x_0-x_1-x_2)\over\sqrt{|\lambda^2|}}
\Psi(x_0,x_1,x_2)\Big\}\;,
\label{l-tran4}
\end{eqnarray}
which is equivalent to the identity Eq.(\ref{l-tran3}).
Here, $\varepsilon=2-{D/2}$ with $D$ denoting the dimension of space-time,
$\Lambda$ is a energy scale to define $x_i=m_i^2/\Lambda^2$
and $x_{_R}=\Lambda_{_{\rm RE}}^2/\Lambda^2$. Additionally, $\lambda^2=x^2+y^2+z^2-2xy-2xz-2yz$,
and the concrete expression of $\Psi(x,y,z)$ can be found in the appendix.
Actually, the equation Eq.(\ref{l-tran4}) provides a crosscheck of Eq.(\ref{2l-vacuum})
and Eq.(\ref{phi}) rather than a verification of Eq.(\ref{l-tran3}).
After applying those identities derived from loop momentum translating invariance,
we formulate the sum of amplitude from those triangle diagrams corresponding
to the self energy Fig.\ref{fig1}(a) satisfying QED gauge invariance and
CPT symmetry explicitly, and extract the Wilson coefficients of those
operators in Eq.(\ref{eq4}).

%%%%%%%%%%%%%%%%%%%%%%%%%%%END MODIFICATION%%%%%%%%%%%%%%%%%%%%%%%%%%%%%

Integrating over loop momenta, one gets the following terms
in the effective Lagrangian:
\begin{eqnarray}
&&{\cal L}^{eff}_{_W}={\sqrt{2}G_{_F}\alpha_{_e}x_{_{\rm w}}\over\pi s_{_{\rm w}}^2Q_{_d}}
V_{_{ts}}^*V_{_{tb}}(4\pi x_{_{\rm R}})^{2\varepsilon}{\Gamma^2(1+\varepsilon)\over
(1-\varepsilon)^2}\Bigg\{\Big(\zeta^{L*}_{_{\alpha\beta}}
\zeta^L_{_{\alpha\beta}}+\zeta^{R*}_{_{\alpha\beta}}\zeta^R_{_{\alpha\beta}}\Big)
\nonumber\\
&&\hspace{1.4cm}\times
\Bigg[{1\over24\varepsilon}\Big\{-\psi_1
+(x_{_{F_\alpha}}+x_{_{F_\beta}})\psi_2\Big\}(x_{_{\rm w}},x_{_t})
-{1\over24}\varrho_{_{2,1}}(x_{_{F_\alpha}},x_{_{F_\beta}})\psi_2(x_{_{\rm w}},x_{_t})
\nonumber\\
&&\hspace{1.4cm}
-{x_{_{F_\alpha}}+x_{_{F_\beta}}\over144}\psi_3(x_{_{\rm w}},x_{_t})
+\phi_1(x_{_{F_\alpha}},x_{_{F_\beta}})\psi_1(x_{_{\rm w}},x_{_t})
+\psi_4(x_{_{\rm w}},x_{_t})
\nonumber\\
&&\hspace{1.4cm}
+F_{_1}(x_{_{\rm w}},x_{_t},x_{_{F_\alpha}},x_{_{F_\beta}})
+Q_{_u}\Bigg({1\over24\varepsilon}\Big\{\psi_5+(x_{_{F_\alpha}}+x_{_{F_\beta}})
\psi_6\Big\}(x_{_{\rm w}},x_{_t})
\nonumber\\
&&\hspace{1.4cm}
-{1\over24}\varrho_{_{2,1}}(x_{_{F_\alpha}},x_{_{F_\beta}})\psi_6(x_{_{\rm w}},x_{_t})
+{x_{_{F_\alpha}}+x_{_{F_\beta}}\over144}\psi_7(x_{_{\rm w}},x_{_t})
\nonumber\\
&&\hspace{1.4cm}
+{1\over8}\phi_2(x_{_{F_\alpha}},x_{_{F_\beta}})\psi_5(x_{_{\rm w}},x_{_t})
+\psi_8(x_{_{\rm w}},x_{_t})
+F_{_2}(x_{_{\rm w}},x_{_t},x_{_{F_\alpha}},x_{_{F_\beta}})
\Bigg)\Bigg]{\cal O}_{_2}
\nonumber\\
%---------------------------------------------------------------------
&&\hspace{1.4cm}
+\Big(\zeta^{L*}_{_{\alpha\beta}}\zeta^L_{_{\alpha\beta}}
-\zeta^{R*}_{_{\alpha\beta}}\zeta^R_{_{\alpha\beta}}\Big)
F_{_3}(x_{_{\rm w}},x_{_t},x_{_{F_\alpha}},x_{_{F_\beta}}){\cal O}_{_2}
\nonumber\\
%---------------------------------------------------------------------
&&\hspace{1.4cm}
+\Big(\zeta^{L*}_{_{\alpha\beta}}\zeta^R_{_{\alpha\beta}}+\zeta^{R*}_{_{\alpha\beta}}
\zeta^L_{_{\alpha\beta}}\Big)(x_{_{F_\alpha}}x_{_{F_\beta}})^{1/2}\Bigg[
-{1\over12\varepsilon}\psi_2(x_{_{\rm w}},x_{_t})
+{1\over12}\varrho_{_{1,1}}(x_{_{F_\alpha}},x_{_{F_\beta}})
\psi_2(x_{_{\rm w}},x_{_t})
\nonumber\\
&&\hspace{1.4cm}
+\phi_3(x_{_{F_\alpha}},x_{_{F_\beta}})\psi_1(x_{_{\rm w}},x_{_t})
+\psi_9(x_{_{\rm w}},x_{_t})+F_{_4}(x_{_{\rm w}},x_{_t},x_{_{F_\alpha}},x_{_{F_\beta}})
\nonumber\\
&&\hspace{1.4cm}
+Q_{_u}\Bigg(-{1\over12\varepsilon}\psi_6(x_{_{\rm w}},x_{_t})
+{1\over12}\varrho_{_{1,1}}(x_{_{F_\alpha}},x_{_{F_\beta}})
\psi_6(x_{_{\rm w}},x_{_t})
+\phi_4(x_{_{F_\alpha}},x_{_{F_\beta}})\psi_5(x_{_{\rm w}},x_{_t})
\nonumber\\
&&\hspace{1.4cm}
+\psi_{10}(x_{_{\rm w}},x_{_t})
+F_{_5}(x_{_{\rm w}},x_{_t},x_{_{F_\alpha}},x_{_{F_\beta}})
\Bigg)\Bigg]{\cal O}_{_2}
\nonumber\\
%---------------------------------------------------------------------
&&\hspace{1.4cm}
+\Big(\zeta^L_{_{\alpha\beta}}\zeta^{R*}_{_{\alpha\beta}}
-\zeta^{L*}_{_{\alpha\beta}}\zeta^R_{_{\alpha\beta}}\Big)(x_{_{F_\alpha}}x_{_{F_\beta}})^{1/2}
F_{_6}(x_{_{\rm w}},x_{_t},x_{_{F_\alpha}},x_{_{F_\beta}}){\cal O}_{_9}
\nonumber\\
%---------------------------------------------------------------------
&&\hspace{1.4cm}
+\Big(\zeta^{L*}_{_{\alpha\beta}}\zeta^L_{_{\alpha\beta}}
+\zeta^{R*}_{_{\alpha\beta}}\zeta^R_{_{\alpha\beta}}\Big)
\Bigg[{1\over24\varepsilon}\Big\{\psi_5
+(x_{_{F_\alpha}}+x_{_{F_\beta}})\psi_6\Big\}(x_{_{\rm w}},x_{_t})
\nonumber\\
&&\hspace{1.4cm}
-{1\over24}\varrho_{_{2,1}}(x_{_{F_\alpha}},x_{_{F_\beta}})\psi_6(x_{_{\rm w}},x_{_t})
+{x_{_{F_\alpha}}+x_{_{F_\beta}}\over144}\psi_7(x_{_{\rm w}},x_{_t})
+{1\over8}\phi_2(x_{_{F_\alpha}},x_{_{F_\beta}})\psi_5(x_{_{\rm w}},x_{_t})
\nonumber\\
&&\hspace{1.4cm}
+\psi_8(x_{_{\rm w}},x_{_t})
+F_{_2}(x_{_{\rm w}},x_{_t},x_{_{F_\alpha}},x_{_{F_\beta}})
+T^c_{_\alpha}F_{_7}(x_{_{\rm w}},x_{_t},x_{_{F_\alpha}},x_{_{F_\beta}})
\Bigg]{\cal O}_{_6}
\nonumber\\
%---------------------------------------------------------------------
&&\hspace{1.4cm}
+\Big(\zeta^{L*}_{_{\alpha\beta}}\zeta^R_{_{\alpha\beta}}+\zeta^{R*}_{_{\alpha\beta}}
\zeta^L_{_{\alpha\beta}}\Big)(x_{_{F_\alpha}}x_{_{F_\beta}})^{1/2}\Bigg[
-{1\over12\varepsilon}\psi_6(x_{_{\rm w}},x_{_t})
+{1\over12}\varrho_{_{1,1}}(x_{_{F_\alpha}},x_{_{F_\beta}})
\psi_6(x_{_{\rm w}},x_{_t})
\nonumber\\
&&\hspace{1.4cm}
+\phi_4(x_{_{F_\alpha}},x_{_{F_\beta}})\psi_5(x_{_{\rm w}},x_{_t})
+\psi_{10}(x_{_{\rm w}},x_{_t})
+F_5(x_{_{\rm w}},x_{_t},x_{_{F_\alpha}},x_{_{F_\beta}})
\nonumber\\
&&\hspace{1.4cm}
+T^c_{_\alpha}F_{_8}(x_{_{\rm w}},x_{_t},x_{_{F_\alpha}},x_{_{F_\beta}})
\Bigg]{\cal O}_{_6}
\nonumber\\
%---------------------------------------------------------------------
&&\hspace{1.4cm}
+\Big(\zeta^{L*}_{_{\alpha\beta}}\zeta^L_{_{\alpha\beta}}
-\zeta^{R*}_{_{\alpha\beta}}\zeta^R_{_{\alpha\beta}}\Big)
T^c_{_\alpha}F_{_9}(x_{_{\rm w}},x_{_t},x_{_{F_\alpha}},x_{_{F_\beta}}){\cal O}_{_6}
\nonumber\\
%---------------------------------------------------------------------
&&\hspace{1.4cm}
+\Big(\zeta^{L*}_{_{\alpha\beta}}\zeta^R_{_{\alpha\beta}}-\zeta^{R*}_{_{\alpha\beta}}
\zeta^L_{_{\alpha\beta}}\Big)(x_{_{F_\alpha}}x_{_{F_\beta}})^{1/2}
T^c_{_\alpha}F_{_{10}}(x_{_{\rm w}},x_{_t},x_{_{F_\alpha}},x_{_{F_\beta}}){\cal O}_{_{12}}
\Bigg\}+\cdots\;,
\label{eff-WFF}
\end{eqnarray}
where $\alpha_{_e}=e^2/4\pi$ and $Q_{_d}=-1/3,\;Q_{_u}=2/3$ represent
the charge of down- and up-type quarks, respectively. $T^c_{_\alpha}=1$
when the heavy virtual fermions take part in the strong interaction, otherwise
$T^c_{_\alpha}=0$. The functions $\psi_i,\;\phi_i$ are defined as
\begin{eqnarray}
%%%%%%%%%%%%%%%%%%%%%%%%%%%%%%%%%%%%%%%%%%%%%%%%%%%%%%%%%%%%%%%%%%%%%%%%%%%%%%%%%%%
&&\psi_1(x,y)={\partial^4\varrho_{_{4,1}}\over\partial x^4}(x,y)
-3{\partial^3\varrho_{_{3,1}}\over\partial x^3}(x,y)\;,
\nonumber\\
%%%%%%%%%%%%%%%%%%%%%%%%%%%%%%%%%%%%%%%%%%%%%%%%%%%%%%%%%%%%%%%%%%%%%%%%%%%%%%%%%%%
&&\psi_2(x,y)={\partial^4\varrho_{_{3,1}}\over\partial x^4}(x,y)
+3{\partial^3\varrho_{_{2,1}}\over\partial x^3}(x,y)\;,
\nonumber\\
%%%%%%%%%%%%%%%%%%%%%%%%%%%%%%%%%%%%%%%%%%%%%%%%%%%%%%%%%%%%%%%%%%%%%%%%%%%%%%%%%%%
&&\psi_3(x,y)=\Big\{4{\partial^4\varrho_{_{3,1}}\over\partial x^4}
-18{\partial^3\varrho_{_{2,1}}\over\partial x^3}
+3{\partial^4\varrho_{_{3,2}}\over\partial x^4}
+9{\partial^3\varrho_{_{2,2}}\over\partial x^3}\Big\}(x,y)\;,
\nonumber\\
%%%%%%%%%%%%%%%%%%%%%%%%%%%%%%%%%%%%%%%%%%%%%%%%%%%%%%%%%%%%%%%%%%%%%%%%%%%%%%%%%%%
&&\psi_4(x,y)=\Big\{{1\over48}{\partial^4\varrho_{_{4,1}}\over\partial x^4}
-{23\over144}{\partial^3\varrho_{_{3,1}}\over\partial x^3}
+{1\over4}{\partial^2\varrho_{_{2,1}}\over\partial x^2}
+{1\over48}{\partial^4\varrho_{_{4,2}}\over\partial x^4}
-{1\over16}{\partial^3\varrho_{_{3,2}}\over\partial x^3}\Big\}(x,y)\;,
\nonumber\\
%%%%%%%%%%%%%%%%%%%%%%%%%%%%%%%%%%%%%%%%%%%%%%%%%%%%%%%%%%%%%%%%%%%%%%%%%%%%%%%%%%%
&&\psi_5(x,y)={\partial^4\varrho_{_{4,1}}\over\partial x^4}(x,y)
-6{\partial^3\varrho_{_{3,1}}\over\partial x^3}(x,y)
+6{\partial^2\varrho_{_{2,1}}\over\partial x^2}(x,y)\;,
\nonumber\\
%%%%%%%%%%%%%%%%%%%%%%%%%%%%%%%%%%%%%%%%%%%%%%%%%%%%%%%%%%%%%%%%%%%%%%%%%%%%%%%%%%%
&&\psi_6(x,y)=6{\partial^2\varrho_{_{1,1}}\over\partial x^2}(x,y)
-{\partial^4\varrho_{_{3,1}}\over\partial x^4}(x,y)\;,
\nonumber\\
%%%%%%%%%%%%%%%%%%%%%%%%%%%%%%%%%%%%%%%%%%%%%%%%%%%%%%%%%%%%%%%%%%%%%%%%%%%%%%%%%%%
&&\psi_7(x,y)=\Big\{4{\partial^4\varrho_{_{3,1}}\over\partial x^4}
-36{\partial^3\varrho_{_{2,1}}\over\partial x^3}
+18{\partial^2\varrho_{_{1,1}}\over\partial x^2}
+3{\partial^4\varrho_{_{3,2}}\over\partial x^4}
-18{\partial^2\varrho_{_{1,2}}\over\partial x^2}\Big\}(x,y)\;,
\nonumber\\
%%%%%%%%%%%%%%%%%%%%%%%%%%%%%%%%%%%%%%%%%%%%%%%%%%%%%%%%%%%%%%%%%%%%%%%%%%%%%%%%%%%
&&\psi_8(x,y)=\Big\{-{1\over48}{\partial^4\varrho_{_{4,1}}\over\partial x^4}
+{19\over72}{\partial^3\varrho_{_{3,1}}\over\partial x^3}
-{2\over3}{\partial^2\varrho_{_{2,1}}\over\partial x^2}
+{1\over3}{\partial\varrho_{_{1,1}}\over\partial x}
-{1\over48}{\partial^4\varrho_{_{4,2}}\over\partial x^4}
\nonumber\\
&&\hspace{2.0cm}
+{1\over8}{\partial^3\varrho_{_{3,2}}\over\partial x^3}
-{1\over8}{\partial^2\varrho_{_{2,2}}\over\partial x^2}\Big\}(x,y)\;,
\nonumber\\
%%%%%%%%%%%%%%%%%%%%%%%%%%%%%%%%%%%%%%%%%%%%%%%%%%%%%%%%%%%%%%%%%%%%%%%%%%%%%%%%%%%
&&\psi_9(x,y)=
\Big\{{1\over72}{\partial^4\varrho_{_{3,1}}\over\partial x^4}
-{3\over8}{\partial^3\varrho_{_{2,1}}\over\partial x^3}
+{1\over24}{\partial^4\varrho_{_{3,2}}\over\partial x^4}
+{1\over8}{\partial^3\varrho_{_{2,2}}\over\partial x^3}\Big\}(x,y)\;,
\nonumber\\
%%%%%%%%%%%%%%%%%%%%%%%%%%%%%%%%%%%%%%%%%%%%%%%%%%%%%%%%%%%%%%%%%%%%%%%%%%%%%%%%%%%
&&\psi_{10}(x,y)=\Big\{-{1\over72}{\partial^4\varrho_{_{3,1}}\over\partial x^4}
+{1\over2}{\partial^3\varrho_{_{2,1}}\over\partial x^3}
-{1\over2}{\partial^2\varrho_{_{1,1}}\over\partial x^2}
-{1\over24}{\partial^4\varrho_{_{3,2}}\over\partial x^4}
+{1\over4}{\partial^2\varrho_{_{1,2}}\over\partial x^2}\Big\}(x,y)\;,
\nonumber\\
%%%%%%%%%%%%%%%%%%%%%%%%%%%%%%%%%%%%%%%%%%%%%%%%%%%%%%%%%%%%%%%%%%%%%%%%%%%%%%%%%%%
&&\phi_1(x,y)=\Big\{{1\over8}{\partial\varrho_{_{2,1}}\over\partial x}
-{1\over24}{\partial^2\varrho_{_{3,1}}\over\partial x^2}
-{3x_{_{\rm w}}\over32}{\partial^2\varrho_{_{2,1}}\over\partial x^2}
+{x_{_{\rm w}}\over16}{\partial^3\varrho_{_{3,1}}\over\partial x^3}
-{x_{_{\rm w}}\over128}{\partial^4\varrho_{_{4,1}}\over\partial x^4}
\Big\}(x,y)\;,
\nonumber\\
%%%%%%%%%%%%%%%%%%%%%%%%%%%%%%%%%%%%%%%%%%%%%%%%%%%%%%%%%%%%%%%%%%%%%%%%%%%%%%%%%%%
&&\phi_2(x,y)=\Big\{-{\partial\varrho_{_{2,1}}\over\partial x}
+{1\over3}{\partial^2\varrho_{_{3,1}}\over\partial x^2}
+{3x_{_{\rm w}}\over4}{\partial^2\varrho_{_{2,1}}\over\partial x^2}
-{x_{_{\rm w}}\over2}{\partial^3\varrho_{_{3,1}}\over\partial x^3}
+{x_{_{\rm w}}\over16}{\partial^4\varrho_{_{4,1}}\over\partial x^4}\Big\}(x,y)\;,
\nonumber\\
%%%%%%%%%%%%%%%%%%%%%%%%%%%%%%%%%%%%%%%%%%%%%%%%%%%%%%%%%%%%%%%%%%%%%%%%%%%%%%%%%%%
&&\phi_3(x,y)=\Big\{{1\over16}{\partial^2\varrho_{_{2,1}}\over\partial x^2}
-{1\over8}{\partial\varrho_{_{1,1}}\over\partial x}
+{x_{_{\rm w}}\over16}{\partial^2\varrho_{_{1,1}}\over\partial x^2}
-{x_{_{\rm w}}\over16}{\partial^3\varrho_{_{2,1}}\over\partial x^3}
+{x_{_{\rm w}}\over96}{\partial^4\varrho_{_{3,1}}\over
\partial x^4}\Big\}(x,y)\;,
\nonumber\\
%%%%%%%%%%%%%%%%%%%%%%%%%%%%%%%%%%%%%%%%%%%%%%%%%%%%%%%%%%%%%%%%%%%%%%%%%%%%%%%%%%%
&&\phi_4(x,y)=\Big\{-{1\over16}{\partial^2\varrho_{_{2,1}}\over\partial x^2}
+{1\over8}{\partial\varrho_{_{1,1}}\over\partial x}
-{x_{_{\rm w}}\over16}{\partial^2\varrho_{_{1,1}}\over\partial x^2}
+{x_{_{\rm w}}\over16}{\partial^3\varrho_{_{2,1}}\over\partial x^3}
-{x_{_{\rm w}}\over96}{\partial^4\varrho_{_{3,1}}\over\partial x^4}\Big\}(x,y)\;.
%%%%%%%%%%%%%%%%%%%%%%%%%%%%%%%%%%%%%%%%%%%%%%%%%%%%%%%%%%%%%%%%%%%%%%%%%%%%%%%%%%%
\label{psi-phi-fun}
\end{eqnarray}
Note that the result in Eq.\ref{eff-WFF} does not depend on the concrete choice
of energy scale $\Lambda$, and the concrete expressions of
$F_i(x,y,z,u),\;\varrho_{_{i,j}}(x,y)\;(i,\;j=1,\;2\;\cdots)$ can be found in appendix.

The charged gauge boson self energy composed of a closed heavy fermion loop
induces the ultraviolet divergence in the Wilson coefficients of effective
Lagrangian, the unrenormalized $W^\pm$ self energy is generally expressed as
\begin{eqnarray}
%%%%%%%%%%%%%%%%%%%%%%%%%%%%%%%%%%%%%%%%%%%%%%%%%%%%%%%%%%%%%%%%%%%%%%%%%%%%%%%%%%%
&&\Sigma_{_{\mu\nu}}^{\rm W}(p,\Lambda_{_{\rm RE}})=\Lambda^2A_0^{\rm w}g_{\mu\nu}+\Big(A_1^{\rm w}
+{p^2\over\Lambda^2}A_2^{\rm w}+\cdots\Big)(p^2g_{\mu\nu}-p_\mu p_\nu)
\nonumber\\
&&\hspace{2.5cm}
+\Big(B_1^{\rm w}+{p^2\over\Lambda^2}B_2^{\rm w}+\cdots\Big)p_\mu p_\nu\;,
%%%%%%%%%%%%%%%%%%%%%%%%%%%%%%%%%%%%%%%%%%%%%%%%%%%%%%%%%%%%%%%%%%%%%%%%%%%%%%%%%%%
\label{eq-w1}
\end{eqnarray}
where the form factors $A_{0,1,2}^{\rm w}$ and $B_{1,2}^{\rm w}$ only depend on
the virtual field masses and renormalization scale.
Here, we omit those terms which are strongly suppressed at the limit
of heavy virtual fermion masses. The corresponding counter terms are given as
\begin{eqnarray}
%%%%%%%%%%%%%%%%%%%%%%%%%%%%%%%%%%%%%%%%%%%%%%%%%%%%%%%%%%%%%%%%%%%%%%%%%%%%%%%%%%%
&&\Sigma_{_{\mu\nu}}^{\rm WC}(p,\Lambda_{_{\rm RE}})=-\Big[\delta m_{_{\rm w}}^2(\Lambda_{_{\rm RE}})
+m_{_{\rm w}}^2\delta Z_{_{\rm w}}(\Lambda_{_{\rm RE}})\Big]g_{\mu\nu}
-\delta Z_{_{\rm w}}(\Lambda_{_{\rm RE}})\Big[p^2g_{\mu\nu}-p_\mu p_\nu\Big]\;.
%%%%%%%%%%%%%%%%%%%%%%%%%%%%%%%%%%%%%%%%%%%%%%%%%%%%%%%%%%%%%%%%%%%%%%%%%%%%%%%%%%%
\label{eq-w2}
\end{eqnarray}

The renormalized self energy is given by
\begin{eqnarray}
%%%%%%%%%%%%%%%%%%%%%%%%%%%%%%%%%%%%%%%%%%%%%%%%%%%%%%%%%%%%%%%%%%%%%%%%%%%%%%%%%%%
&&\hat{\Sigma}_{_{\mu\nu}}^{\rm W}(p,\Lambda_{_{\rm RE}})=
\Sigma_{_{\mu\nu}}^{\rm W}(p,\Lambda_{_{\rm RE}})
+\Sigma_{_{\mu\nu}}^{\rm WC}(p,\Lambda_{_{\rm RE}})\;.
%%%%%%%%%%%%%%%%%%%%%%%%%%%%%%%%%%%%%%%%%%%%%%%%%%%%%%%%%%%%%%%%%%%%%%%%%%%%%%%%%%%
\label{eq-w3}
\end{eqnarray}
For on-shell external gauge boson $W^\pm$, we have \cite{onshell}
\begin{eqnarray}
%%%%%%%%%%%%%%%%%%%%%%%%%%%%%%%%%%%%%%%%%%%%%%%%%%%%%%%%%%%%%%%%%%%%%%%%%%%%%%%%%%%
&&\hat{\Sigma}_{_{\mu\nu}}^{\rm W}(p,m_{_{\rm w}})\epsilon^\nu(p)\Big|_{p^2=m_{_{\rm w}}^2}=0
\;,\nonumber\\
&&\lim\limits_{p^2\rightarrow m_{_{\rm w}}^2}{1\over p^2-m_{_{\rm w}}^2}
\hat{\Sigma}_{_{\mu\nu}}^{\rm W}(p,m_{_{\rm w}})\epsilon^\nu(p)=\epsilon_{_\mu}(p)\;,
%%%%%%%%%%%%%%%%%%%%%%%%%%%%%%%%%%%%%%%%%%%%%%%%%%%%%%%%%%%%%%%%%%%%%%%%%%%%%%%%%%%
\label{eq-w4}
\end{eqnarray}
where $\epsilon(p)$ is the polarization vector of $W^\pm$ gauge boson.
Inserting Eq. (\ref{eq-w1}) and Eq. (\ref{eq-w2}) into Eq. (\ref{eq-w4}),
we derive the counter terms for the $W^\pm$ self energy in on-shell scheme as
\begin{eqnarray}
%%%%%%%%%%%%%%%%%%%%%%%%%%%%%%%%%%%%%%%%%%%%%%%%%%%%%%%%%%%%%%%%%%%%%%%%%%%%%%%%%%%
&&\delta Z_{_{\rm w}}^{os}=A_1^{\rm w}+{m_{_{\rm w}}^2\over\Lambda^2}A_2^{\rm w}
=A_1^{\rm w}+x_{_{\rm z}}A_2^{\rm w}\;,
\nonumber\\
&&\delta m_{_{\rm w}}^{2,os}=A_0^{\rm w}\Lambda^2
-m_{_{\rm w}}^2\delta Z_{_{\rm w}}^{os}\;.
%%%%%%%%%%%%%%%%%%%%%%%%%%%%%%%%%%%%%%%%%%%%%%%%%%%%%%%%%%%%%%%%%%%%%%%%%%%%%%%%%%%
\label{eq-w5}
\end{eqnarray}
To cancel the ultraviolet divergence and those dangerous terms violating the
decoupling theorem completely, we should derive the counter term for the vertex
$\gamma W^+W^-$ here since the corresponding coupling is not zero at tree level.
In the nonlinear $R_\xi$ gauge with $\xi=1$,
the counter term for the vertex $\gamma W^+W^-$ is
\begin{eqnarray}
%%%%%%%%%%%%%%%%%%%%%%%%%%%%%%%%%%%%%%%%%%%%%%%%%%%%%%%%%%%%%%%%%%%%%%%%%%%%%%%%%%%
&&i\delta C_{\gamma W^+W^-}=ie\cdot\delta Z_{_{\rm w}}(\Lambda_{_{\rm RE}})
\Big[g_{\mu\nu}(k_1-k_2)_\rho+g_{\nu\rho}(k_2-k_3)_\mu+g_{\rho\mu}(k_3-k_1)_\nu\Big]\;,
%%%%%%%%%%%%%%%%%%%%%%%%%%%%%%%%%%%%%%%%%%%%%%%%%%%%%%%%%%%%%%%%%%%%%%%%%%%%%%%%%%%
\label{eq-w6}
\end{eqnarray}
where $k_i\;(i=1,\;2,\;3)$ denote the incoming momenta of $W^\pm$ and photon,
and $\mu,\;\nu,\;\rho$ denote the corresponding Lorentz indices respectively.

We can verify that the sum of amplitude from counter diagrams
satisfies the Ward identity required by the QED gauge invariance obviously.
Accordingly, the effective Lagrangian from the counter term diagrams is written as
\begin{eqnarray}
%%%%%%%%%%%%%%%%%%%%%%%%%%%%%%%%%%%%%%%%%%%%%%%%%%%%%%%%%%%%%%%%%%%%%%
&&\delta{\cal L}_{_{W}}^C=
-{\sqrt{2}G_{_F}\alpha_{_e}x_{_{\rm w}}\over\pi s_{_{\rm w}}^2Q_{_d}}
V_{_{ts}}^*V_{_{tb}}(4\pi x_{_{\rm R}})^{2\varepsilon}{\Gamma^2(1+\varepsilon)\over
(1-\varepsilon)^2}\Bigg\{\Big(\zeta^{L*}_{_{\alpha\beta}}
\zeta^L_{_{\alpha\beta}}+\zeta^{R*}_{_{\alpha\beta}}\zeta^R_{_{\alpha\beta}}\Big)
\nonumber\\
&&\hspace{1.4cm}\times
\Bigg[{1\over24\varepsilon}\Big\{-\psi_1
+(x_{_{F_\alpha}}+x_{_{F_\beta}})\psi_2\Big\}(x_{_{\rm w}},x_{_t})
-{1\over24}\varrho_{_{2,1}}(x_{_{F_\alpha}},x_{_{F_\beta}})\psi_2(x_{_{\rm w}},x_{_t})
\nonumber\\
&&\hspace{1.4cm}
-{x_{_{F_\alpha}}+x_{_{F_\beta}}\over144}\psi_3(x_{_{\rm w}},x_{_t})
+\phi_1(x_{_{F_\alpha}},x_{_{F_\beta}})\psi_1(x_{_{\rm w}},x_{_t})
+\psi_4(x_{_{\rm w}},x_{_t})
\nonumber\\
&&\hspace{1.4cm}
+Q_{_u}\Bigg({1\over24\varepsilon}\Big\{\psi_5+(x_{_{F_\alpha}}+x_{_{F_\beta}})
\psi_6\Big\}(x_{_{\rm w}},x_{_t})
-{1\over24}\varrho_{_{2,1}}(x_{_{F_\alpha}},x_{_{F_\beta}})\psi_6(x_{_{\rm w}},x_{_t})
\nonumber\\
&&\hspace{1.4cm}
+{x_{_{F_\alpha}}+x_{_{F_\beta}}\over144}\psi_7(x_{_{\rm w}},x_{_t})
+{1\over8}\phi_2(x_{_{F_\alpha}},x_{_{F_\beta}})\psi_5(x_{_{\rm w}},x_{_t})
+\psi_8(x_{_{\rm w}},x_{_t})\Bigg)\Bigg]{\cal O}_{_2}
\nonumber\\
%---------------------------------------------------------------------
&&\hspace{1.4cm}
+\Big(\zeta^{L*}_{_{\alpha\beta}}\zeta^R_{_{\alpha\beta}}+\zeta^{R*}_{_{\alpha\beta}}
\zeta^L_{_{\alpha\beta}}\Big)(x_{_{F_\alpha}}x_{_{F_\beta}})^{1/2}\Bigg[
-{1\over12\varepsilon}\psi_2(x_{_{\rm w}},x_{_t})
\nonumber\\
&&\hspace{1.4cm}
+{1\over12}\varrho_{_{1,1}}(x_{_{F_\alpha}},x_{_{F_\beta}})
\psi_2(x_{_{\rm w}},x_{_t})
+\phi_3(x_{_{F_\alpha}},x_{_{F_\beta}})\psi_1(x_{_{\rm w}},x_{_t})
+\psi_9(x_{_{\rm w}},x_{_t})
\nonumber\\
&&\hspace{1.4cm}
+Q_{_u}\Bigg(-{1\over12\varepsilon}\psi_6(x_{_{\rm w}},x_{_t})
+{1\over12}\varrho_{_{1,1}}(x_{_{F_\alpha}},x_{_{F_\beta}})
\psi_6(x_{_{\rm w}},x_{_t})
\nonumber\\
&&\hspace{1.4cm}
+\phi_4(x_{_{F_\alpha}},x_{_{F_\beta}})\psi_5(x_{_{\rm w}},x_{_t})
+\psi_{10}(x_{_{\rm w}},x_{_t})\Bigg)\Bigg]{\cal O}_{_2}
\nonumber\\
%---------------------------------------------------------------------
&&\hspace{1.4cm}
+\Big(\zeta^{L*}_{_{\alpha\beta}}\zeta^L_{_{\alpha\beta}}
+\zeta^{R*}_{_{\alpha\beta}}\zeta^R_{_{\alpha\beta}}\Big)
\Bigg[{1\over24\varepsilon}\Big\{\psi_5
+(x_{_{F_\alpha}}+x_{_{F_\beta}})\psi_6\Big\}(x_{_{\rm w}},x_{_t})
+{x_{_{F_\alpha}}+x_{_{F_\beta}}\over144}\psi_7(x_{_{\rm w}},x_{_t})
\nonumber\\
&&\hspace{1.4cm}
-{1\over24}\varrho_{_{2,1}}(x_{_{F_\alpha}},x_{_{F_\beta}})\psi_6(x_{_{\rm w}},x_{_t})
+{1\over8}\phi_2(x_{_{F_\alpha}},x_{_{F_\beta}})\psi_5(x_{_{\rm w}},x_{_t})
+\psi_8(x_{_{\rm w}},x_{_t})\Bigg]{\cal O}_{_6}
\nonumber\\
%---------------------------------------------------------------------
&&\hspace{1.4cm}
+\Big(\zeta^{L*}_{_{\alpha\beta}}\zeta^R_{_{\alpha\beta}}+\zeta^{R*}_{_{\alpha\beta}}
\zeta^L_{_{\alpha\beta}}\Big)(x_{_{F_\alpha}}x_{_{F_\beta}})^{1/2}\Bigg[
-{1\over12\varepsilon}\psi_6(x_{_{\rm w}},x_{_t})
+{1\over12}\varrho_{_{1,1}}(x_{_{F_\alpha}},x_{_{F_\beta}})
\psi_6(x_{_{\rm w}},x_{_t})
\nonumber\\
&&\hspace{1.4cm}
+\phi_4(x_{_{F_\alpha}},x_{_{F_\beta}})\psi_5(x_{_{\rm w}},x_{_t})
+\psi_{10}(x_{_{\rm w}},x_{_t})\Bigg]{\cal O}_{_6}\Bigg\}+\cdots\;.
%%%%%%%%%%%%%%%%%%%%%%%%%%%%%%%%%%%%%%%%%%%%%%%%%%%%%%%%%%%%%%%%%%%%%%%%%%%%%%%%%%%
\label{w-counter}
\end{eqnarray}

%%%%%%%%%%%%%%%%%%%%%%%%%BEGIN MODIFICATION%%%%%%%%%%%%%%%%%%%%%%%%%%%%%
Adding the counter terms to bare Lagrangian Eq.(\ref{eff-WFF}),
we cancel the ultraviolet divergence there. Under our approximation,
the resulted effective Lagrangian is written as
\begin{eqnarray}
%%%%%%%%%%%%%%%%%%%%%%%%%%%%%%%%%%%%%%%%%%%%%%%%%%%%%%%%%%%%%%%%%%%%%%
&&\widehat{\cal L}^{eff}_{_W}={\sqrt{2}G_{_F}\alpha_{_e}x_{_{\rm
w}}\over\pi s_{_{\rm
w}}^2Q_{_d}}V_{_{ts}}^*V_{_{tb}}\Bigg\{\Big(\zeta^{L*}_{_{\alpha\beta}}
\zeta^L_{_{\alpha\beta}}+\zeta^{R*}_{_{\alpha\beta}}\zeta^R_{_{\alpha\beta}}\Big)
\Bigg[F_{_1} +Q_{_u}F_{_2} \Bigg](x_{_{\rm
w}},x_{_t},x_{_{F_\alpha}},x_{_{F_\beta}}){\cal O}_{_2}
\nonumber\\
%---------------------------------------------------------------------
&&\hspace{1.4cm}
+\Big(\zeta^{L*}_{_{\alpha\beta}}\zeta^L_{_{\alpha\beta}}
-\zeta^{R*}_{_{\alpha\beta}}\zeta^R_{_{\alpha\beta}}\Big)
F_{_3}(x_{_{\rm w}},x_{_t},x_{_{F_\alpha}},x_{_{F_\beta}}){\cal O}_{_2}
\nonumber\\
%---------------------------------------------------------------------
&&\hspace{1.4cm}
+\Big(\zeta^{L*}_{_{\alpha\beta}}\zeta^R_{_{\alpha\beta}}+\zeta^{R*}_{_{\alpha\beta}}
\zeta^L_{_{\alpha\beta}}\Big)(x_{_{F_\alpha}}x_{_{F_\beta}})^{1/2}\Bigg[
F_{_4} +Q_{_u}F_{_5}\Bigg](x_{_{\rm
w}},x_{_t},x_{_{F_\alpha}},x_{_{F_\beta}}) {\cal O}_{_2}
\nonumber\\
%---------------------------------------------------------------------
&&\hspace{1.4cm}
+\Big(\zeta^L_{_{\alpha\beta}}\zeta^{R*}_{_{\alpha\beta}}
-\zeta^{L*}_{_{\alpha\beta}}\zeta^R_{_{\alpha\beta}}\Big)(x_{_{F_\alpha}}x_{_{F_\beta}})^{1/2}
F_{_6}(x_{_{\rm w}},x_{_t},x_{_{F_\alpha}},x_{_{F_\beta}}){\cal O}_{_9}
\nonumber\\
%---------------------------------------------------------------------
&&\hspace{1.4cm}
+\Big(\zeta^{L*}_{_{\alpha\beta}}\zeta^L_{_{\alpha\beta}}
+\zeta^{R*}_{_{\alpha\beta}}\zeta^R_{_{\alpha\beta}}\Big)
\Bigg[F_{_2} +T^c_{_\alpha}F_{_7}\Bigg](x_{_{\rm
w}},x_{_t},x_{_{F_\alpha}},x_{_{F_\beta}}) {\cal O}_{_6}
\nonumber\\
%---------------------------------------------------------------------
&&\hspace{1.4cm}
+\Big(\zeta^{L*}_{_{\alpha\beta}}\zeta^R_{_{\alpha\beta}}+\zeta^{R*}_{_{\alpha\beta}}
\zeta^L_{_{\alpha\beta}}\Big)(x_{_{F_\alpha}}x_{_{F_\beta}})^{1/2}\Bigg[
F_5 +T^c_{_\alpha}F_{_8}\Bigg](x_{_{\rm
w}},x_{_t},x_{_{F_\alpha}},x_{_{F_\beta}}) {\cal O}_{_6}
\nonumber\\
%---------------------------------------------------------------------
&&\hspace{1.4cm}
+\Big(\zeta^{L*}_{_{\alpha\beta}}\zeta^L_{_{\alpha\beta}}
-\zeta^{R*}_{_{\alpha\beta}}\zeta^R_{_{\alpha\beta}}\Big)
T^c_{_\alpha}F_{_9}(x_{_{\rm w}},x_{_t},x_{_{F_\alpha}},x_{_{F_\beta}}){\cal O}_{_6}
\nonumber\\
%---------------------------------------------------------------------
&&\hspace{1.4cm}
+\Big(\zeta^{L*}_{_{\alpha\beta}}\zeta^R_{_{\alpha\beta}}-\zeta^{R*}_{_{\alpha\beta}}
\zeta^L_{_{\alpha\beta}}\Big)(x_{_{F_\alpha}}x_{_{F_\beta}})^{1/2}
T^c_{_\alpha}F_{_{10}}(x_{_{\rm w}},x_{_t},x_{_{F_\alpha}},x_{_{F_\beta}}){\cal O}_{_{12}}
\Bigg\}+\cdots\;,
\label{renor-eff-WFF}
\end{eqnarray}
which only depends on the masses of virtual fields. It should be
clarified that the corrections to the coefficients of ${\cal O}_{_{9,12}}$
do not depend on the concrete renormalization scheme adopted here since the relevant
terms from bare Lagrangian do not contain the ultraviolet divergence.
In the limit $z\ll x,y$, the function $\Phi(x,y,z)$ can be approximated
in powers of $z$ as
\begin{eqnarray}
%%%%%%%%%%%%%%%%%%%%%%%%%%%%%%%%%%%%%%%%%%%%%%%%%%%%%%%%%%%%%%%%%%%%%%
&&\Phi(x,y,z)=\varphi_0(x,y)+z\varphi_1(x,y)+{z^2\over2!}\varphi_2(x,y)
+{z^3\over3!}\varphi_3(x,y)
\nonumber\\
&&\hspace{2.2cm}
+2z\Big(\ln z-1\Big)\pi_{_1}(x,y)
+2z^2\Big({\ln z\over2!}-{3\over4}\Big)\pi_{_2}(x,y)
\nonumber\\
&&\hspace{2.2cm}
+2z^3\Big({\ln z\over3!}-{11\over36}\Big)\pi_{_3}(x,y)+\cdots
%%%%%%%%%%%%%%%%%%%%%%%%%%%%%%%%%%%%%%%%%%%%%%%%%%%%%%%%%%%%%%%%%%%%%%
\label{phi-expand}
\end{eqnarray}
with
\begin{eqnarray}
%%%%%%%%%%%%%%%%%%%%%%%%%%%%%%%%%%%%%%%%%%%%%%%%%%%%%%%%%%%%%%%%%%%%%%
&&\pi_{_1}(x,y)=1+\varrho_{_{1,1}}(x,y),
\nonumber\\
&&\pi_{_2}(x,y)=-{x+y\over(x-y)^2}-{2xy\over(x-y)^3}\ln{y\over x},
\nonumber\\
&&\pi_{_3}(x,y)=-{1\over(x-y)^2}-{12xy\over(x-y)^4}
-{6xy(x+y)\over(x-y)^5}\ln{y\over x}\;,
%%%%%%%%%%%%%%%%%%%%%%%%%%%%%%%%%%%%%%%%%%%%%%%%%%%%%%%%%%%%%%%%%%%%%%
\label{pi1}
\end{eqnarray}
and the concrete expressions of function $\varphi_i(x,y)\;(i=0,1,2,3)$
can be found in appendix. Using the asymptotic expressions in Eq.(\ref{phi-expand}),
we derive the leading contributions contained in
Eq.\ref{renor-eff-WFF} under the assumption $m_{_F}=m_{_{F_\alpha}}=m_{_{F_\beta}}
\gg m_{_{\rm w}}$:
\begin{eqnarray}
%%%%%%%%%%%%%%%%%%%%%%%%%%%%%%%%%%%%%%%%%%%%%%%%%%%%%%%%%%%%%%%%%%%%%%
&&\widehat{\cal L}^{eff}_{_W}\approx
{\sqrt{2}G_{_F}\alpha_{_e}x_{_{\rm w}}\over\pi s_{_{\rm w}}^2Q_{_d}}
V_{_{ts}}^*V_{_{tb}}\Bigg\{\Big(\zeta^{L*}_{_{\alpha\beta}}
\zeta^L_{_{\alpha\beta}}+\zeta^{R*}_{_{\alpha\beta}}\zeta^R_{_{\alpha\beta}}\Big)
\nonumber\\
&&\hspace{1.4cm}\times
\Bigg[\Big\{-{1-3Q_\beta\over8}{\partial^2\varrho_{_{2,1}}\over\partial x_{_{\rm w}}^2}
-{2-3Q_\beta\over8}{\partial\varrho_{_{1,1}}\over\partial x_{_{\rm w}}}
-{1\over144}{\partial^4\varrho_{_{4,1}}\over\partial x_{_{\rm w}}^4}
-{1\over48}{\partial^3\varrho_{_{3,1}}\over\partial x_{_{\rm w}}^3}\Big\}(x_{_{\rm w}},x_{_t})
\nonumber\\
&&\hspace{1.4cm}
+Q_{_u}\Big\{{1\over144}{\partial^4\varrho_{_{4,1}}\over\partial x_{_{\rm w}}^4}
-{1\over12}{\partial^3\varrho_{_{3,1}}\over\partial x_{_{\rm w}}^3}
-{29\over72}{\partial^2\varrho_{_{2,1}}\over\partial x_{_{\rm w}}^2}
-{11\over12}{\partial\varrho_{_{1,1}}\over\partial x_{_{\rm w}}}\Big\}
(x_{_{\rm w}},x_{_t})\Bigg]{\cal O}_{_2}
\nonumber\\
%---------------------------------------------------------------------
&&\hspace{1.4cm}
-{1-Q_\beta\over8}\Big(\zeta^{L*}_{_{\alpha\beta}}\zeta^L_{_{\alpha\beta}}
-\zeta^{R*}_{_{\alpha\beta}}\zeta^R_{_{\alpha\beta}}\Big)
{\partial\varrho_{_{1,1}}\over\partial x_{_{\rm w}}}(x_{_{\rm w}},x_{_t}){\cal O}_{_2}
\nonumber\\
%---------------------------------------------------------------------
&&\hspace{1.4cm}
+\Big(\zeta^{L*}_{_{\alpha\beta}}\zeta^R_{_{\alpha\beta}}+\zeta^{R*}_{_{\alpha\beta}}
\zeta^L_{_{\alpha\beta}}\Big)\Bigg[
\Big\{{1\over144}{\partial^4\varrho_{_{4,1}}\over\partial x_{_{\rm w}}^4}
%\nonumber\\
%%---------------------------------------------------------------------
%&&\hspace{2.8cm}
-{1\over16}{\partial^3\varrho_{_{3,1}}\over\partial x_{_{\rm w}}^3}
+{1\over4}{\partial^2\varrho_{_{2,1}}\over\partial x_{_{\rm w}}^2}
+{1\over16}{\partial\varrho_{_{1,1}}\over\partial x_{_{\rm w}}}\Big\}
(x_{_{\rm w}},x_{_t})
\nonumber\\
&&\hspace{1.4cm}
+Q_{_u}\Big\{-{1\over144}{\partial^4\varrho_{_{4,1}}\over\partial x_{_{\rm w}}^4}
+{1\over12}{\partial^3\varrho_{_{3,1}}\over\partial x_{_{\rm w}}^3}
-{5\over24}{\partial^2\varrho_{_{2,1}}\over\partial x_{_{\rm w}}^2}
-{1\over12}{\partial\varrho_{_{1,1}}\over\partial x_{_{\rm w}}}\Big\}
(x_{_{\rm w}},x_{_t})\Bigg]{\cal O}_{_2}
\nonumber\\
%---------------------------------------------------------------------
&&\hspace{1.4cm}
+{1\over8}\Big(\zeta^L_{_{\alpha\beta}}\zeta^{R*}_{_{\alpha\beta}}
-\zeta^{L*}_{_{\alpha\beta}}\zeta^R_{_{\alpha\beta}}\Big)
{\partial^2\varrho_{_{2,1}}\over\partial x_{_{\rm w}}^2}
(x_{_{\rm w}},x_{_t}){\cal O}_{_9}
\nonumber\\
%---------------------------------------------------------------------
&&\hspace{1.4cm}
+\Big(\zeta^{L*}_{_{\alpha\beta}}\zeta^L_{_{\alpha\beta}}
+\zeta^{R*}_{_{\alpha\beta}}\zeta^R_{_{\alpha\beta}}\Big)
\Bigg[\Big\{{1\over144}{\partial^4\varrho_{_{4,1}}\over\partial x_{_{\rm w}}^4}
-{1\over12}{\partial^3\varrho_{_{3,1}}\over\partial x_{_{\rm w}}^3}
-{29\over72}{\partial^2\varrho_{_{2,1}}\over\partial x_{_{\rm w}}^2}
-{11\over12}{\partial\varrho_{_{1,1}}\over\partial x_{_{\rm w}}}\Big\}
(x_{_{\rm w}},x_{_t})
\nonumber\\
&&\hspace{1.4cm}
+T^c_{_\alpha}\Big\{{3\over8}
{\partial^2\varrho_{_{2,1}}\over\partial x^2}(x,y)
+{5\over4}{\partial\varrho_{_{1,1}}\over\partial x}\Big\}
(x_{_{\rm w}},x_{_t})\Bigg]{\cal O}_{_6}
\nonumber\\
%---------------------------------------------------------------------
&&\hspace{1.4cm}
+\Big(\zeta^{L*}_{_{\alpha\beta}}\zeta^R_{_{\alpha\beta}}+\zeta^{R*}_{_{\alpha\beta}}
\zeta^L_{_{\alpha\beta}}\Big)
\Big\{-{1\over144}{\partial^4\varrho_{_{4,1}}\over\partial x_{_{\rm w}}^4}
+{1\over12}{\partial^3\varrho_{_{3,1}}\over\partial x_{_{\rm w}}^3}
-{5\over24}{\partial^2\varrho_{_{2,1}}\over\partial x_{_{\rm w}}^2}
\nonumber\\
&&\hspace{1.4cm}
-{1\over12}{\partial\varrho_{_{1,1}}\over\partial x_{_{\rm w}}}\Big\}
(x_{_{\rm w}},x_{_t}){\cal O}_{_6}
\nonumber\\
%---------------------------------------------------------------------
&&\hspace{1.4cm}
+\Big(\zeta^{L*}_{_{\alpha\beta}}\zeta^R_{_{\alpha\beta}}-\zeta^{R*}_{_{\alpha\beta}}
\zeta^L_{_{\alpha\beta}}\Big)
T^c_{_\alpha}\Big\{{1\over16}{\partial\varrho_{_{1,1}}\over\partial x_{_{\rm w}}}
+{7\over24}{\partial^2\varrho_{_{2,1}}\over\partial x_{_{\rm w}}^2}\Big\}
(x_{_{\rm w}},x_{_t}){\cal O}_{_{12}}
\Bigg\}+\cdots\;,
%%%%%%%%%%%%%%%%%%%%%%%%%%%%%%%%%%%%%%%%%%%%%%%%%%%%%%%%%%%%%%%%%%%%%%
\label{asyHF-lag-W}
\end{eqnarray}
where ellipses represent those relatively unimportant corrections.
Comparing the result in Eq.(\ref{renor-eff-WFF}), the contributions from
the corresponding diagrams contain
the additional suppressed factor $m_{_b}^2/\Lambda_{_{\rm EW}}^2$
when both of virtual charged gauge bosons in Fig.\ref{fig1}(a)
are replaced with the charged Goldstone $G^\pm$. However, we should consider
the corrections from those two loop diagrams in which one of virtual
charged gauge bosons is replaced with the charged Goldstone $G^\pm$ since it
represents the longitudinal component of charged gauge boson
in nonlinear $R_\xi$ gauge. As the closed fermion loop is attached
to virtual $W^\pm$ gauge boson and charged Higgs simultaneously, the corresponding
triangle diagrams belong to the famous Barr-Zee type diagrams \cite{Barr-Zee}.
It is shown \cite{Pilaftsis} that this type diagrams contribute to important
corrections to the effective Lagrangian. For the reason mentioned above,
we also generalize the result directly to the diagrams
in which a closed heavy loop is attached to the virtual $H^\pm$
and $W^\pm$ fields simultaneously.

\subsection{The corrections from the diagrams where a closed heavy
fermion loop is attached to the virtual $W^\pm,\;G^\pm\;(H^\pm)$ bosons}
\indent\indent
Similarly, the renormalizable interaction among the EW charged
Goldstone/Higgs $G^\pm\;(H^\pm)$ and the heavy fermions $F_{\alpha,\beta}$
can be expressed in a more universal form as
\begin{eqnarray}
&&{\cal L}_{_{S^\pm FF}}={e\over s_{_{\rm w}}}\Big[G^{-}\bar{F}_\alpha
({\cal G}^{c,L}_{_{\alpha\beta}}\omega_-+{\cal G}^{c,R}_{_{\alpha\beta}}\omega_+)F_\beta
+H^{-}\bar{F}_\alpha({\cal H}^{c,L}_{_{\alpha\beta}}\omega_-
+{\cal H}^{c,R}_{_{\alpha\beta}}\omega_+)F_\beta\Big]+h.c.\;,
\label{charged-G-H-FF}
\end{eqnarray}
where the concrete expressions of ${\cal G}^{c,L,R}_{_{\alpha\beta}},\;
{\cal H}^{c,L,R}_{_{\alpha\beta}}$ depend on the
models employed in our calculation, the conservation of electric charge
requires $Q_\beta-Q_\alpha=1$.
Generally, the couplings among the charged Goldstone/Higgs
and quarks are written as
\begin{eqnarray}
&&{\cal L}_{_{S^\pm\bar{d}u}}={eV_{_{ud}}^*\over\sqrt{2}
s_{_{\rm w}}}\Big\{G^{-}\bar{d}\Big[{m_{_u}\over m_{_{\rm w}}}\omega_+
+{m_{_d}\over m_{_{\rm w}}}\omega_-\Big]u
+H^{-}\bar{d}\Big[{m_{_u}\over m_{_{\rm w}}}\omega_+
-{\cal B}_{c}{m_{_d}\over m_{_{\rm w}}}\omega_-\Big]u\Big\}+h.c.\;,
\label{charged-G-H-lepton}
\end{eqnarray}
where the parameter ${\cal B}_{c}$ also depends on the concrete
models adopted in our analysis. In full theory,
the couplings in Eq.(\ref{charged-G-H-FF}) induce the corrections
to the effective Lagrangian for $b\rightarrow s\gamma$ through the
diagrams presented in Fig.\ref{fig1}(b, c).

Since there is no mixing between the charged gauge boson and charged
Higgs/Goldstone at tree level, the corresponding corrections from the diagrams presented
in Fig.\ref{fig1}(b, c) to the bare effective Lagrangian do not include
the ultraviolet divergence, and can be formulated as
\begin{eqnarray}
%%%%%%%%%%%%%%%%%%%%%%%%%%%%%%%%%%%%%%%%%%%%%%%%%%%%%%%%%%%%%%%%%%%%%%
&&\widehat{\cal L}^{eff}_{_{WH}}={\sqrt{2}G_{_F}\alpha_{_e}{\cal B}_{_c}
\over\pi s_{_{\rm w}}^2Q_{_d}}V_{_{ts}}^*V_{_{tb}}
\Bigg\{(x_{_{F_\beta}}x_{_{\rm w}})^{1/2}
P_1(x_{_{\rm w}},x_{_{H^\pm}},x_{_t},x_{_{F_\alpha}},x_{_{F_\beta}})
\nonumber\\
%---------------------------------------------------------------------
&&\hspace{1.2cm}\times
\Big[\Re\Big({\cal H}^{c,L}_{_{\beta\alpha}}\zeta^L_{_{\alpha\beta}}
+{\cal H}^{c,R}_{_{\beta\alpha}}\zeta^R_{_{\alpha\beta}}\Big)
{\cal O}_{_5}
-i\Im\Big({\cal H}^{c,L}_{_{\beta\alpha}}\zeta^L_{_{\alpha\beta}}
+{\cal H}^{c,R}_{_{\beta\alpha}}\zeta^R_{_{\alpha\beta}}\Big)
{\cal O}_{_{11}}\Big]
\nonumber\\
%---------------------------------------------------------------------
&&\hspace{1.2cm}
+(x_{_{\rm w}}x_{_{F_\alpha}})^{1/2}
P_2(x_{_{\rm w}},x_{_{H^\pm}},x_{_t},x_{_{F_\alpha}},x_{_{F_\beta}})
\Big[\Re\Big({\cal H}^{c,L}_{_{\beta\alpha}}\zeta^R_{_{\alpha\beta}}
+{\cal H}^{c,R}_{_{\beta\alpha}}\zeta^L_{_{\alpha\beta}}\Big){\cal O}_{_5}
\nonumber\\
&&\hspace{1.2cm}
-i\Im\Big({\cal H}^{c,L}_{_{\beta\alpha}}\zeta^R_{_{\alpha\beta}}
+{\cal H}^{c,R}_{_{\beta\alpha}}\zeta^L_{_{\alpha\beta}}\Big){\cal O}_{_{11}}\Big]
\nonumber\\
%---------------------------------------------------------------------
&&\hspace{1.2cm}
+(x_{_{\rm w}}x_{_{F_\beta}})^{1/2}
P_3(x_{_{\rm w}},x_{_{H^\pm}},x_{_t},x_{_{F_\alpha}},x_{_{F_\beta}})
\Big[\Re\Big({\cal H}^{c,L}_{_{\beta\alpha}}\zeta^L_{_{\alpha\beta}}
-{\cal H}^{c,R}_{_{\beta\alpha}}\zeta^R_{_{\alpha\beta}}\Big){\cal O}_{_5}
\nonumber\\
&&\hspace{1.2cm}
-i\Im\Big({\cal H}^{c,L}_{_{\beta\alpha}}\zeta^L_{_{\alpha\beta}}
-{\cal H}^{c,R}_{_{\beta\alpha}}\zeta^R_{_{\alpha\beta}}\Big){\cal O}_{_{11}}\Big]
\nonumber\\
%---------------------------------------------------------------------
&&\hspace{1.2cm}
+(x_{_{\rm w}}x_{_{F_\alpha}})^{1/2}
P_4(x_{_{\rm w}},x_{_{H^\pm}},x_{_t},x_{_{F_\alpha}},x_{_{F_\beta}})
\Big[\Re\Big({\cal H}^{c,L}_{_{\beta\alpha}}\zeta^R_{_{\alpha\beta}}
-{\cal H}^{c,R}_{_{\beta\alpha}}\zeta^L_{_{\alpha\beta}}\Big){\cal O}_{_5}
\nonumber\\
&&\hspace{1.2cm} -i\Im\Big({\cal
H}^{c,L}_{_{\beta\alpha}}\zeta^R_{_{\alpha\beta}} -{\cal
H}^{c,R}_{_{\beta\alpha}}\zeta^L_{_{\alpha\beta}}\Big){\cal
O}_{_{11}}\Big]
\nonumber\\
%---------------------------------------------------------------------
&&\hspace{1.2cm}
+(x_{_{F_\beta}}x_{_{\rm w}})^{1/2}P_5
(x_{_{\rm w}},x_{_{H^\pm}},x_{_t},x_{_{F_\alpha}},x_{_{F_\beta}})
\Big[\Re\Big({\cal H}^{c,L}_{_{\beta\alpha}}\zeta^L_{_{\alpha\beta}}
+{\cal H}^{c,R}_{_{\beta\alpha}}\zeta^R_{_{\alpha\beta}}\Big)
{\cal O}_{_8}
\nonumber\\
&&\hspace{1.2cm}
-i\Im\Big({\cal H}^{c,L}_{_{\beta\alpha}}\zeta^L_{_{\alpha\beta}}
+{\cal H}^{c,R}_{_{\beta\alpha}}\zeta^R_{_{\alpha\beta}}\Big)
{\cal O}_{_{13}}\Big]
\nonumber\\
%---------------------------------------------------------------------
&&\hspace{1.2cm}
+(x_{_{F_\beta}}x_{_{\rm w}})^{1/2}
P_5(x_{_{\rm w}},x_{_{H^\pm}},x_{_t},x_{_{F_\alpha}},x_{_{F_\beta}})
\Big[\Re\Big({\cal H}^{c,L}_{_{\beta\alpha}}\zeta^R_{_{\alpha\beta}}
+{\cal H}^{c,R}_{_{\beta\alpha}}\zeta^L_{_{\alpha\beta}}\Big)
{\cal O}_{_8}
\nonumber\\
&&\hspace{1.2cm}
+i\Im\Big({\cal H}^{c,L}_{_{\beta\alpha}}\zeta^R_{_{\alpha\beta}}
+{\cal H}^{c,R}_{_{\beta\alpha}}\zeta^L_{_{\alpha\beta}}\Big)
{\cal O}_{_{13}}\Big]
\nonumber\\
%---------------------------------------------------------------------
&&\hspace{1.2cm}
+(x_{_{F_\beta}}x_{_{\rm w}})^{1/2}P_6
(x_{_{\rm w}},x_{_{H^\pm}},x_{_t},x_{_{F_\alpha}},x_{_{F_\beta}})
\Big[\Re\Big({\cal H}^{c,L}_{_{\beta\alpha}}\zeta^L_{_{\alpha\beta}}
-{\cal H}^{c,R}_{_{\beta\alpha}}\zeta^R_{_{\alpha\beta}}\Big)
{\cal O}_{_8}
\nonumber\\
&&\hspace{1.2cm}
-i\Im\Big({\cal H}^{c,L}_{_{\beta\alpha}}\zeta^L_{_{\alpha\beta}}
-{\cal H}^{c,R}_{_{\beta\alpha}}\zeta^R_{_{\alpha\beta}}\Big)
{\cal O}_{_{13}}\Big]
\nonumber\\
%---------------------------------------------------------------------
&&\hspace{1.2cm}
-(x_{_{F_\beta}}x_{_{\rm w}})^{1/2}
P_6(x_{_{\rm w}},x_{_{H^\pm}},x_{_t},x_{_{F_\alpha}},x_{_{F_\beta}})
\Big[\Re\Big({\cal H}^{c,L}_{_{\beta\alpha}}\zeta^R_{_{\alpha\beta}}
+{\cal H}^{c,R}_{_{\beta\alpha}}\zeta^L_{_{\alpha\beta}}\Big)
{\cal O}_{_8}
\nonumber\\
&&\hspace{1.2cm}
+i\Im\Big({\cal H}^{c,L}_{_{\beta\alpha}}\zeta^R_{_{\alpha\beta}}
+{\cal H}^{c,R}_{_{\beta\alpha}}\zeta^L_{_{\alpha\beta}}\Big)
{\cal O}_{_{13}}\Big]\Bigg\}\;,
\nonumber\\
%%%%%%%%%%%%%%%%%%%%%%%%%%%%%%%%%%%%%%%%%%%%%%%%%%%%%%%%%%%%%%%%%%%%%%
&&\widehat{\cal L}^{eff}_{_{WG}}=\widehat{\cal L}^{eff}_{_{WH}}
({\cal B}_{c}\rightarrow1,
{\cal G}^{c,L,R}_{_{\beta\alpha}}\rightarrow{\cal H}^{c,L,R}_{_{\beta\alpha}}
,x_{_{H^\pm}}\rightarrow x_{_{\rm w}})\;.
%%%%%%%%%%%%%%%%%%%%%%%%%%%%%%%%%%%%%%%%%%%%%%%%%%%%%%%%%%%%%%%%%%%%%%
\label{MED-W-G-H}
\end{eqnarray}
The expressions of form factors $P_i(x,y,z,u,w)\;(i=1,\cdots,4)$ can
be found in appendix.

Using the asymptotic expressions of $\Phi(x,y,z)$
at the limit $x,\;y\gg z$ in Eq.\ref{phi-expand},
we simplify the expressions of Eq.(\ref{MED-W-G-H})
in the limit $m_{_F}=m_{_{F_\alpha}}=m_{_{F_\beta}}\gg m_{_{\rm w}}$ as:
\begin{eqnarray}
%%%%%%%%%%%%%%%%%%%%%%%%%%%%%%%%%%%%%%%%%%%%%%%%%%%%%%%%%%%%%%%%%%%%%%
%%%%%%%%%%%%%%%%%%%%%%%%%%%%%%%%%%%%%%%%%%%%%%%%%%%%%%%%%%%%%%%%%%%%%%
&&\widehat{\cal L}^{eff}_{_{WH}}\approx{\sqrt{2}G_{_F}\alpha_{_e}{\cal B}_{_c}
m_{_{\rm w}}\over\pi s_{_{\rm w}}^2Q_{_d}m_{_F}}
V_{_{ts}}^*V_{_{tb}}\Bigg\{\Big[{21\over64}-{5\over288}Q_\beta
\nonumber\\
%---------------------------------------------------------------------
&&\hspace{1.2cm}
+({3\over16}+{Q_\beta\over48})\Big(\ln m_{_F}^2
-{\varrho_{_{2,1}}(m_{_{\rm w}}^2,m_{_t}^2)-\varrho_{_{2,1}}(m_{_{H^\pm}}^2,m_{_t}^2)
\over m_{_{\rm w}}^2-m_{_{H^\pm}}^2}\Big)\Big]
\nonumber\\
&&\hspace{1.2cm}\times
\Big[\Re\Big({\cal H}^{c,L}_{_{\beta\alpha}}\zeta^L_{_{\alpha\beta}}
+{\cal H}^{c,R}_{_{\beta\alpha}}\zeta^R_{_{\alpha\beta}}\Big)
{\cal O}_{_5}
-i\Im\Big({\cal H}^{c,L}_{_{\beta\alpha}}\zeta^L_{_{\alpha\beta}}
+{\cal H}^{c,R}_{_{\beta\alpha}}\zeta^R_{_{\alpha\beta}}\Big)
{\cal O}_{_{11}}\Big]
\nonumber\\
%---------------------------------------------------------------------
&&\hspace{1.2cm}
+\Big[{19-20Q_\beta\over144}+{2-4Q_\beta\over48}\Big(\ln m_{_F}^2
-{\varrho_{_{2,1}}(m_{_{\rm w}}^2,m_{_t}^2)-\varrho_{_{2,1}}(m_{_{H^\pm}}^2,m_{_t}^2)
\over m_{_{\rm w}}^2-m_{_{H^\pm}}^2}\Big)\Big]
\nonumber\\
&&\hspace{1.2cm}\times
\Big[\Re\Big({\cal H}^{c,L}_{_{\beta\alpha}}\zeta^R_{_{\alpha\beta}}
+{\cal H}^{c,R}_{_{\beta\alpha}}\zeta^L_{_{\alpha\beta}}\Big){\cal O}_{_5}
-i\Im\Big({\cal H}^{c,L}_{_{\beta\alpha}}\zeta^R_{_{\alpha\beta}}
+{\cal H}^{c,R}_{_{\beta\alpha}}\zeta^L_{_{\alpha\beta}}\Big){\cal O}_{_{11}}\Big]
\nonumber\\
%---------------------------------------------------------------------
&&\hspace{1.2cm}
-\Big[{16\over144}+{2+6Q_\beta\over48}\Big(\ln m_{_F}^2
-{\varrho_{_{2,1}}(m_{_{\rm w}}^2,m_{_t}^2)-\varrho_{_{2,1}}(m_{_{H^\pm}}^2,m_{_t}^2)
\over m_{_{\rm w}}^2-m_{_{H^\pm}}^2}\Big)\Big]
\nonumber\\
&&\hspace{1.2cm}\times
\Big[\Re\Big({\cal H}^{c,L}_{_{\beta\alpha}}\zeta^L_{_{\alpha\beta}}
-{\cal H}^{c,R}_{_{\beta\alpha}}\zeta^R_{_{\alpha\beta}}\Big){\cal O}_{_5}
-i\Im\Big({\cal H}^{c,L}_{_{\beta\alpha}}\zeta^L_{_{\alpha\beta}}
-{\cal H}^{c,R}_{_{\beta\alpha}}\zeta^R_{_{\alpha\beta}}\Big){\cal O}_{_{11}}\Big]
\nonumber\\
%---------------------------------------------------------------------
&&\hspace{1.2cm}
-\Big[{2Q_\beta\over144}+{6-2Q_\beta\over48}\Big(\ln m_{_F}^2
-{\varrho_{_{2,1}}(m_{_{\rm w}}^2,m_{_t}^2)-\varrho_{_{2,1}}(m_{_{H^\pm}}^2,m_{_t}^2)
\over m_{_{\rm w}}^2-m_{_{H^\pm}}^2}\Big)\Big]
\nonumber\\
&&\hspace{1.2cm}\times \Big[\Re\Big({\cal
H}^{c,L}_{_{\beta\alpha}}\zeta^R_{_{\alpha\beta}} -{\cal
H}^{c,R}_{_{\beta\alpha}}\zeta^L_{_{\alpha\beta}}\Big){\cal O}_{_5}
-i\Im\Big({\cal H}^{c,L}_{_{\beta\alpha}}\zeta^R_{_{\alpha\beta}}
-{\cal H}^{c,R}_{_{\beta\alpha}}\zeta^L_{_{\alpha\beta}}\Big){\cal
O}_{_{11}}\Big]
\nonumber\\
%---------------------------------------------------------------------
&&\hspace{1.2cm}
-{1\over8\sqrt{2}}\Big[1+\ln m_{_F}^2
-{\varrho_{_{2,1}}(m_{_{\rm w}}^2,m_{_t}^2)-\varrho_{_{2,1}}(m_{_{H^\pm}}^2,m_{_t}^2)
\over m_{_{\rm w}}^2-m_{_{H^\pm}}^2}\Big]
\nonumber\\
&&\hspace{1.2cm}\times
\Big[\Re\Big({\cal H}^{c,L}_{_{\beta\alpha}}\zeta^L_{_{\alpha\beta}}
+{\cal H}^{c,R}_{_{\beta\alpha}}\zeta^R_{_{\alpha\beta}}\Big){\cal O}_{_8}
-i\Im\Big({\cal H}^{c,L}_{_{\beta\alpha}}\zeta^L_{_{\alpha\beta}}
+{\cal H}^{c,R}_{_{\beta\alpha}}\zeta^R_{_{\alpha\beta}}\Big)
{\cal O}_{_{13}}\Big]
\nonumber\\
%---------------------------------------------------------------------
&&\hspace{1.2cm}
-{1\over8\sqrt{2}}\Big[1+\ln m_{_F}^2
-{\varrho_{_{2,1}}(m_{_{\rm w}}^2,m_{_t}^2)-\varrho_{_{2,1}}(m_{_{H^\pm}}^2,m_{_t}^2)
\over m_{_{\rm w}}^2-m_{_{H^\pm}}^2}\Big]
\nonumber\\
&&\hspace{1.2cm}\times
\Big[\Re\Big({\cal H}^{c,L}_{_{\beta\alpha}}\zeta^R_{_{\alpha\beta}}
+{\cal H}^{c,R}_{_{\beta\alpha}}\zeta^L_{_{\alpha\beta}}\Big){\cal O}_{_8}
+i\Im\Big({\cal H}^{c,L}_{_{\beta\alpha}}\zeta^R_{_{\alpha\beta}}
+{\cal H}^{c,R}_{_{\beta\alpha}}\zeta^L_{_{\alpha\beta}}\Big)
{\cal O}_{_{13}}\Big]
\nonumber\\
%---------------------------------------------------------------------
&&\hspace{1.2cm}
-{1\over4\sqrt{2}}\Big[1+\ln m_{_F}^2
-{\varrho_{_{2,1}}(m_{_{\rm w}}^2,m_{_t}^2)-\varrho_{_{2,1}}(m_{_{H^\pm}}^2,m_{_t}^2)
\over m_{_{\rm w}}^2-m_{_{H^\pm}}^2}\Big]
\nonumber\\
&&\hspace{1.2cm}\times
\Big[\Re\Big({\cal H}^{c,L}_{_{\beta\alpha}}\zeta^L_{_{\alpha\beta}}
-{\cal H}^{c,R}_{_{\beta\alpha}}\zeta^R_{_{\alpha\beta}}\Big){\cal O}_{_8}
-i\Im\Big({\cal H}^{c,L}_{_{\beta\alpha}}\zeta^L_{_{\alpha\beta}}
-{\cal H}^{c,R}_{_{\beta\alpha}}\zeta^R_{_{\alpha\beta}}\Big)
{\cal O}_{_{13}}\Big]
\nonumber\\
%---------------------------------------------------------------------
&&\hspace{1.2cm}
+{1\over4\sqrt{2}}\Big[1+\ln m_{_F}^2
-{\varrho_{_{2,1}}(m_{_{\rm w}}^2,m_{_t}^2)-\varrho_{_{2,1}}(m_{_{H^\pm}}^2,m_{_t}^2)
\over m_{_{\rm w}}^2-m_{_{H^\pm}}^2}\Big]
\nonumber\\
&&\hspace{1.2cm}\times
\Big[\Re\Big({\cal H}^{c,L}_{_{\beta\alpha}}\zeta^R_{_{\alpha\beta}}
+{\cal H}^{c,R}_{_{\beta\alpha}}\zeta^L_{_{\alpha\beta}}\Big){\cal O}_{_8}
+i\Im\Big({\cal H}^{c,L}_{_{\beta\alpha}}\zeta^R_{_{\alpha\beta}}
+{\cal H}^{c,R}_{_{\beta\alpha}}\zeta^L_{_{\alpha\beta}}\Big)
{\cal O}_{_{13}}\Big]\Bigg\}\;.
%%%%%%%%%%%%%%%%%%%%%%%%%%%%%%%%%%%%%%%%%%%%%%%%%%%%%%%%%%%%%%%%%%%%%%
\label{ASY-MED-W-GH}
\end{eqnarray}
The results indicate that the corrections to the effective Lagrangian from
the diagrams presented in Fig.\ref{fig1}(b, c) are suppressed in the limit
$m_{_F}=m_{_{F_\alpha}}=m_{_{F_\beta}}\gg m_{_{\rm w}}$
unless the couplings ${\cal H}^{c,L,R}_{_{\beta\alpha}}$ violate the
decoupling theorem.

It is well known that the short distance QCD affects the rare $B$
decay strongly. At the NLO level \cite{QCD-Run},
the Wilson coefficients at the bottom quark scale are given as
\begin{eqnarray}
&&\tilde{C}_5(\mu_ b)\approx0.67(\tilde{C}_5(\mu_{_{\rm w}})
-0.42\tilde{C}_8(\mu_{_{\rm w}})-0.88)\;,
\nonumber\\
&&\tilde{C}_8(\mu_ b)\approx0.7(\tilde{C}_8(\mu_{_{\rm w}})+0.12)\;,
\label{running}
\end{eqnarray}
where the corresponding Wilson coefficients at EW scale
are written as
\begin{eqnarray}
&&\tilde{C}_5(\mu_{_{\rm w}})=C_2(\mu_{_{\rm w}})+C_5(\mu_{_{\rm w}})
+C_9(\mu_{_{\rm w}})+C_{11}(\mu_{_{\rm w}})\;,
\nonumber\\
&&\tilde{C}_8(\mu_{_{\rm w}})=C_6(\mu_{_{\rm w}})+C_8(\mu_{_{\rm w}})
+C_{12}(\mu_{_{\rm w}})+C_{13}(\mu_{_{\rm w}})\;.
\label{re-wilson}
\end{eqnarray}

As an application, we investigate the relative corrections to
the branching ratio of rare decay $B\rightarrow X_{_s}\gamma$
originating from those sectors.

\section{The corrections to branching ratio of $B\rightarrow X_{_s}\gamma$\label{sec3}}
\indent\indent
%%%%%%%%%%%%%%%%%%%%%%%%%%%%%%%%%%%%%%%%%%%%%%%%%%%%%%%%%%%%%%%%%%%
\begin{figure}
\setlength{\unitlength}{1mm}
\begin{center}
\begin{picture}(0,120)(0,0)
\put(-60,-20){\includegraphics{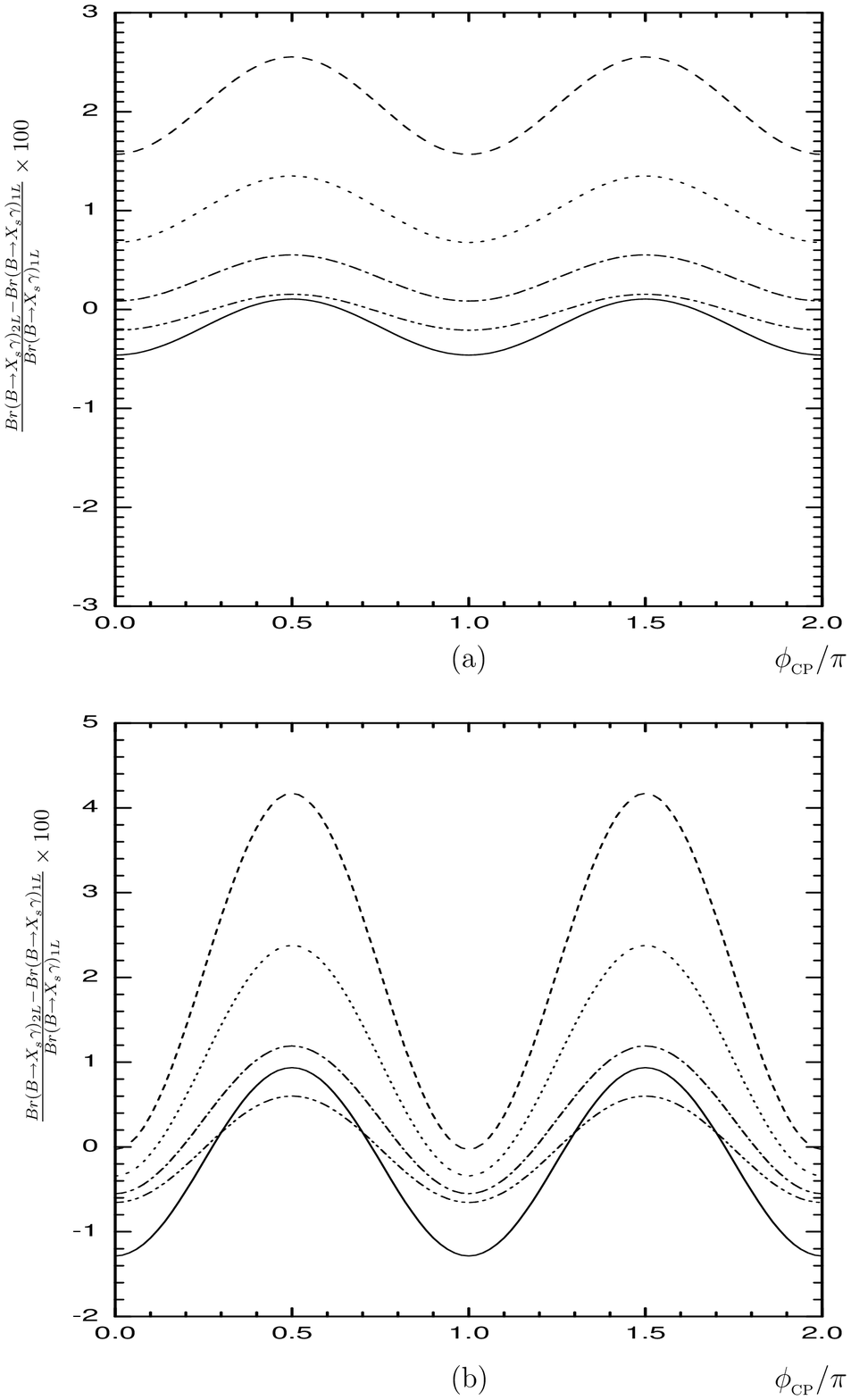}}
\end{picture}
\caption[]{The relative correction to the branching ratio of the
inclusive $B\rightarrow X_{_s}\gamma$ decay versus the possible
CP violation phases $\phi_{_{\rm CP}}$.
Where the solid-line represents the theoretical correction
with $Q_\beta=1,\;Q_\alpha=0$,
the dash-line represents the theoretical correction with
$Q_\beta=-1/3,\;Q_\alpha=-4/3$, the dot-line represents
the theoretical correction with $Q_\beta=2/3,\;Q_\alpha=-1/3$,
the dash-dot-line represents
the theoretical correction with $Q_\beta=4/3,\;Q_\alpha=1/3$,
and the dash-dot-dot-line represents
the theoretical correction with $Q_\beta=5/3,\;Q_\alpha=2/3$,
respectively. The enhancing factor is chosen as the trivial
${\cal B}_c=1$ in (a), or a nontrivial ${\cal B}_c=10$ in (b)}
\label{fig3}
\end{center}
\end{figure}
%%%%%%%%%%%%%%%%%%%%%%%%%%%%%%%%%%%%%%%%%%%%%%%%%%%%%%%%%%%%%%%%%%%
In order to eliminate the strong dependence
on the b-quark mass, the branching ratio is usually normalized by
the decay rate of the $B$ meson semileptonic decay:
\begin{eqnarray}
&&{\Gamma(B\rightarrow X_{_s}\gamma)\over\Gamma(B\rightarrow
X_{_c}e\bar{\nu})}={\Gamma(b-s\gamma)\over\Gamma(b-c\overline{e}\nu)}
={2\alpha_{_e}\over3\pi\rho(y)\chi(y)}|\tilde{C}_5(\mu_b)|^2\;,
\label{eq11}
\end{eqnarray}
where $\rho(y)=1-8y+8y^3-y^4-12y^2\ln y$ is the phase-space factor
with $y=(m_{_c}/m_{_b})^2$, and $\chi(y)=1-\frac{2\alpha_s(m_b)}{3\pi}f(y)$
with $f(m_c^2/m_b^2)\approx2.4$.
From now on we shall assume the value $BR(B\rightarrow
X_{_c}e\bar{\nu})=10.5\%$ for the semileptonic branching ratio,
$\alpha_{_s}(m_{_{\rm z}})=0.118$, $\alpha_{_e}(m_{_{\rm z}})=1/127$. For the mass
spectrum of SM, we take $m_{_t}=174\;{\rm GeV},\;m_{_b}=4.2\;{\rm GeV}
,\;m_{_{\rm w}}=80.42\;{\rm GeV}$ and $m_{_{\rm z}}=91.19\;{\rm GeV}$.
In the CKM matrix, we apply the
Wolfenstein parameterization and set $A=0.85,\;\lambda=0.22,\;
\rho=0.22,\;\eta=0.35$ \cite{Data}.

Without loss of generality, we adopt the universal assumptions
on those couplings and mass spectrum of new physics as
\begin{eqnarray}
&&\zeta^L_{_{\alpha\beta}}=\zeta^R_{_{\alpha\beta}}={\cal H}^{c,L}_{_{\beta\alpha}}
={\cal H}^{c,R}_{_{\beta\alpha}}={\cal G}^{c,L}_{_{\beta\alpha}}={\cal G}^{c,R}_{_{\beta\alpha}}
=e^{i\phi_{_{\rm CP}}}\;,
\nonumber\\
&&m_{_{F_\alpha}}=m_{_{F_\beta}}=m_{_{H^\pm}}=\Lambda_{_{\rm NP}}
\label{assumptions-couplings}
\end{eqnarray}

To continue our discussion, we assume the electric charge of heavy
fermions as $Q_\beta=2/3,\;1,\;-1/3,\;4/3,\;5/3$, which corresponds
to the electric charge of another heavy fermion in inner loop
$Q_\alpha=-1/3,\;0,\;4/3,\;1/3,\;2/3$ respectively. In addition,
we also assume that those heavy fermions with fractional electric charge
all take part in strong interaction.

Many extensions of the SM include the heavy fermion fields with
$Q_\beta=2/3,\;Q_\alpha=-1/3$. In the extensions of SM
with large \cite{L-extD} or warped \cite{W-extD} extra dimensions,
the KK excitations of up- and down-type quarks form a closed
fermion loop which can be attached to
the zero modes of charged gauge boson and Higgs. In the minimal
supersymmetric extension of SM (MSSM) \cite{Su}, the closed fermion loop
composed by chargino ($Q_\beta=1$) and neutralino ($Q_\alpha=0$)
can be attached to the charged gauge boson and Higgs. In the $3-3-1$
model \cite{3-3-1}, the electric charge of exotic quarks are assigned as
$Q_\beta=4/3,\;5/3$.

In many EW extensions of the SM,
the couplings among the charged Higgs and quarks contain an enhancing
factor ${\cal B}_c$. For example, ${\cal B}_c=\tan\beta$ in the MSSM
is a strong enhancing factor at large $\tan\beta$ limit. In other
EW theories such as the littlest Higgs \cite{LHT}, $3-3-1$ model,
the couplings among the charged Higgs and quarks also contain a nontrivial
enhancing factor ${\cal B}_c\gg1$. In our numerical discussion,
we assume the possible enhancing factor with a trivial value ${\cal B}_c=1$
or a nontrivial value ${\cal B}_c=10$.

%%%%%%%%%%%%%%%%%%%%%%%%%%%%%%%%%%%%%%%%%%%%%%%%%%%%%%%%%%%%%%%%%%%
\begin{figure}
\setlength{\unitlength}{1mm}
\begin{center}
\begin{picture}(0,50)(0,0)
\put(-50,-80){\includegraphics{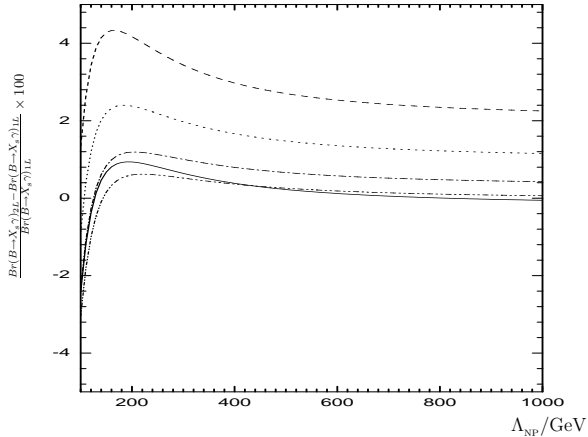}}
\end{picture}
\caption[]{The relative correction to the branching ratio of the
inclusive $B\rightarrow X_{_s}\gamma$ decay versus the energy
scale of new physics $\Lambda_{_{\rm NP}}$.
Where the solid-line represents the theoretical correction
with $Q_\beta=1,\;Q_\alpha=0$,
the dash-line represents the theoretical correction with
$Q_\beta=-1/3,\;Q_\alpha=-4/3$, the dot-line represents
the theoretical correction with $Q_\beta=2/3,\;Q_\alpha=-1/3$,
the dash-dot-line represents
the theoretical correction with $Q_\beta=4/3,\;Q_\alpha=1/3$,
and the dash-dot-dot-line represents
the theoretical correction with $Q_\beta=5/3,\;Q_\alpha=2/3$,
respectively.}
\label{fig4}
\end{center}
\end{figure}
%%%%%%%%%%%%%%%%%%%%%%%%%%%%%%%%%%%%%%%%%%%%%%%%%%%%%%%%%%%%%%%%%%%

Including NLO QCD effects, we plot the relative corrections to one loop
SM theoretical prediction on branching ratio of the inclusive
$B\rightarrow X_{_s}\gamma$ decay versus the possible CP violation
phase $\phi_{_{\rm CP}}$ with ${\cal B}_c=1$ in FIG. \ref{fig3}(a).
Depending on concrete choices of $Q_\beta$ and the CP violation
phase $\phi_{_{\rm CP}}$, the relative corrections to the branching ratio
from those two loop diagrams can reach $2.5\%$. Comparing with the corrections
from QCD, the modifications from those two loop EW diagrams
are unimportant certainly. Nevertheless, those effects can be observed
possibly in the experiment along with improving of the theoretical analysis
and increasing of the experiment precision. Taking the enhancing factor
${\cal B}_c=10$, we plot the relative corrections to one loop
SM theoretical prediction on branching ratio of the inclusive
$B\rightarrow X_{_s}\gamma$ decay versus the possible CP violation
phase $\phi_{_{\rm CP}}$ in FIG. \ref{fig3}(b).
Because the contributions from two loop Bar-Zee diagrams are
enhanced drastically, the relative corrections to one loop
SM theoretical prediction on the branching ratio of
$B\rightarrow X_{_s}\gamma$ can reach $4.5\%$. Although
the two-loop EW corrections can not compete with that from QCD,
we cannot neglecte the corrections with this magnitude.

Assuming ${\cal B}_c=10$ and $\phi_{_{\rm CP}}=\pi/2$, we plot the relative corrections to one loop
SM theoretical prediction on the branching ratio of $B\rightarrow X_{_s}\gamma$
varying with the energy scale of new physics $\Lambda_{_{\rm NP}}$
in FIG. \ref{fig4}. Since the intervention between the top quark and
the particles in new physics, the relative corrections reach the maximum
($\sim4.5\%$) around $\Lambda_{_{\rm NP}}=200\;{\rm GeV}$. With increasing
of $\Lambda_{_{\rm NP}}$, the relative corrections turn smaller and smaller.
At $\Lambda_{_{\rm NP}}=1\;{\rm TeV}$, the relative corrections
are about $2\%$.

In the SM, the $CP$ asymmetry of the $B\rightarrow X_{_s}\gamma$ process
is calculated to be rather small: $A_{_{CP}}\sim 0.5\%$ \cite{Kagan}.
Certainly, the new CP violation phases may induce the observable
effects on the $CP$ asymmetry of $B\rightarrow X_{_s}\gamma$.
However, the numerical results indicate that the corrections from
those two loop diagrams to the $CP$ asymmetry of $B\rightarrow X_{_s}\gamma$
are rather small. The relative correction to one loop
SM theoretical prediction on the branching ratio of $B\rightarrow X_{_s}\gamma$
is already above $8\%$ when ${\cal B}_c=30$, $\Lambda_{_{\rm NP}}=200\;{\rm GeV}$
and $\phi_{_{\rm CP}}=\pi/2$,
the corresponding correction from those two loop diagrams to the $CP$ asymmetry
is still smaller than $1\%$ under our universal assumptions on the parameter
space.

As mentioned above, the universal assumptions on the couplings and mass
spectrum of new physics are adopted in our numerical analysis. In concrete
EW extensions of the SM, this choice is a very simple assumption
on parameter space. However, the numerical results given above
reflect the typical magnitude of corrections from those two loop diagrams to the
branching ratio of $B\rightarrow X_{_s}\gamma$ unless
there is contingent cancelation among different sectors of those two loop diagrams
in concrete extensions of the SM.

\section{Conclusions\label{sec4}}
\indent\indent
Applying effective Lagrangian method and on-shell scheme,
we analyze the EW corrections to the rare decay $b\rightarrow s+\gamma$
from some special two loop diagrams in which a closed
heavy fermion loop is attached to the virtual charged gauge bosons or Higgs.
The analysis shows that the final results satisfy the decoupling theorem explicitly
when the interactions among Higgs and heavy fermions do not contain
the nondecoupling couplings. Adopting the universal assumptions on
the relevant couplings and masses of new physics, we present
the relative corrections from those two loop diagrams to one loop
SM theoretical prediction on the branching ratio of $B\rightarrow X_{_s}\gamma$
varying with the possible CP violation phases and energy scale of
new physics. The numerical results indicate that the relative corrections
from those two loop diagrams can reach $5\%$ if there is not contingent
cancelation among different sectors of corresponding contributions.

\begin{acknowledgments}
The work has been supported by the National Natural Science Foundation of China (NNSFC)
with Grant No. 10675027 and 10975027.
\end{acknowledgments}
\vspace{2.0cm}
\appendix

\section{Form factors in the two-loop Wilson coefficients\label{appb}}
\indent\indent
\indent\indent

The definition of $\Psi(x,y,z)$ is written as:
\begin{itemize}
\item $\lambda^2>0,\;\sqrt{y}+\sqrt{z}<\sqrt{x}$:
\begin{eqnarray}
&&\Psi(x,y,z)=2\ln\Big({x+y-z-\lambda\over2x}\Big)
\ln\Big({x-y+z-\lambda\over2x}\Big)-\ln{y\over x}\ln{z\over x}
\nonumber\\
&&\hspace{2.2cm}
-2L_{i_2}\Big({x+y-z-\lambda\over2x}\Big)
-2L_{i_2}\Big({x-y+z-\lambda\over2x}\Big)+{\pi^2\over3}\;,
\label{aeq2}
\end{eqnarray}
where $L_{i_2}(x)$ is the spence function;

\item $\lambda^2>0,\;\sqrt{x}+\sqrt{z}<\sqrt{y}$:
\begin{eqnarray}
&&\Psi(x,y,z)={\rm Eq.}(\ref{aeq2})(x\leftrightarrow y)\;;
\label{aeq3}
\end{eqnarray}

\item $\lambda^2>0,\;\sqrt{x}+\sqrt{y}<\sqrt{z}$:
\begin{eqnarray}
&&\Psi(x,y,z)={\rm Eq.}(\ref{aeq2})(x\leftrightarrow z)\;;
\label{aeq4}
\end{eqnarray}

\item $\lambda^2<0$:
\begin{eqnarray}
&&\Psi(x,y,z)=2\Big\{Cl_2\Big(2\arccos(
{-x+y+z\over2\sqrt{yz}})\Big)
+Cl_2\Big(2\arccos({x-y+z\over2\sqrt{xz}})\Big)
\nonumber\\
&&\hspace{2.2cm}
+Cl_2\Big(2\arccos({x+y-z\over2\sqrt{xy}})\Big)\Big\}\;,
\label{aeq12}
\end{eqnarray}
where $Cl_2(x)$ denotes the Clausen function.
\end{itemize}

The expressions of $\varphi_0(x,y),\;\varphi_1(x,y),\;\varphi_2(x,y)$
and $\varphi_3(x,y)$ are given as
\begin{eqnarray}
%%%%%%%%%%%%%%%%%%%%%%%%%%%%%%%%%%%%%%%%%%%%%%%%%%%%%%%%%%%%%%%%%%%%%%
&&\varphi_0(x,y)=\left\{\begin{array}{ll}(x+y)\ln x\ln y+(x-y)\Theta(x,y)
\;,&x>y\;;\\
2x\ln^2x\;,&x=y\;;\\
(x+y)\ln x\ln y+(y-x)\Theta(y,x)
\;,&x<y\;.\end{array}\right.
%%%%%%%%%%%%%%%%%%%%%%%%%%%%%%%%%%%%%%%%%%%%%%%%%%%%%%%%%%%%%%%%%%%%%%
\label{varphi0}
\end{eqnarray}

\begin{eqnarray}
%%%%%%%%%%%%%%%%%%%%%%%%%%%%%%%%%%%%%%%%%%%%%%%%%%%%%%%%%%%%%%%%%%%%%%
&&\varphi_1(x,y)=\left\{\begin{array}{ll}-\ln x\ln y-{x+y\over x-y}\Theta(x,y)
\;,&x>y\;;\\
4-2\ln x-\ln^2x\;,&x=y\;;\\
-\ln x\ln y-{x+y\over y-x}\Theta(y,x)
\;,&x<y\;,\end{array}\right.
%%%%%%%%%%%%%%%%%%%%%%%%%%%%%%%%%%%%%%%%%%%%%%%%%%%%%%%%%%%%%%%%%%%%%%
\label{varphi1}
\end{eqnarray}

\begin{eqnarray}
%%%%%%%%%%%%%%%%%%%%%%%%%%%%%%%%%%%%%%%%%%%%%%%%%%%%%%%%%%%%%%%%%%%%%%
&&\varphi_2(x,y)=\left\{\begin{array}{ll}
{(2x^2+6xy)\ln x-(6xy+2y^2)\ln y\over(x-y)^3}-{4xy\over(x-y)^3}\Theta(x,y)
\;,&x>y\;;\\
-{5\over9x}+{2\over3x}\ln x\;,&x=y\;;\\
{(2x^2+6xy)\ln x-(6xy+2y^2)\ln y\over(x-y)^3}-{4xy\over(y-x)^3}\Theta(y,x)
\;,&x<y\;,\end{array}\right.
%%%%%%%%%%%%%%%%%%%%%%%%%%%%%%%%%%%%%%%%%%%%%%%%%%%%%%%%%%%%%%%%%%%%%%
\label{varphi2}
\end{eqnarray}

\begin{eqnarray}
%%%%%%%%%%%%%%%%%%%%%%%%%%%%%%%%%%%%%%%%%%%%%%%%%%%%%%%%%%%%%%%%%%%%%%
&&\varphi_3(x,y)=\left\{\begin{array}{ll}-{12xy(x+y)\over(x-y)^5}\Theta(x,y)
-{2(x^2+6xy+y^2)\over(x-y)^4} & \\
+{2(x^3+20x^2y+11xy^2)\ln x-2(y^3+20xy^2+11x^2y)\ln y\over(x-y)^5}
\;,&x>y\;;\\
-{53\over150x^2}+{1\over5x^2}\ln x\;,&x=y\;;\\
-{12xy(x+y)\over(y-x)^5}\Theta(y,x)-{2(x^2+6xy+y^2)\over(x-y)^4} & \\
+{2(x^3+20x^2y+11xy^2)\ln x-2(y^3+20xy^2+11x^2y)\ln y\over(x-y)^5}
\;,&x<y\;,
\end{array}\right.
%%%%%%%%%%%%%%%%%%%%%%%%%%%%%%%%%%%%%%%%%%%%%%%%%%%%%%%%%%%%%%%%%%%%%%
\label{varphi3}
\end{eqnarray}
with
\begin{eqnarray}
%%%%%%%%%%%%%%%%%%%%%%%%%%%%%%%%%%%%%%%%%%%%%%%%%%%%%%%%%%%%%%%%%%%%%%
&&\Theta(x,y)=\ln x\ln{y\over x}-2\ln(x-y)\ln{y\over x}-2Li_2({y\over x})+{\pi^2\over3}\;.
%%%%%%%%%%%%%%%%%%%%%%%%%%%%%%%%%%%%%%%%%%%%%%%%%%%%%%%%%%%%%%%%%%%%%%
\label{theta}
\end{eqnarray}

The functions adopted in the text are written as
\begin{eqnarray}
%%%%%%%%%%%%%%%%%%%%%%%%%%%%%%%%%%%%%%%%%%%%%%%%%%%%%%%%%%%%%%%%%%%%%%
&&\varrho_{_{i,j}}(x,y)={x^i\ln^jx-y^i\ln^jy\over x-y}\;,\;\;
%\nonumber\\
%%%%%%%%%%%%%%%%%%%%%%%%%%%%%%%%%%%%%%%%%%%%%%%%%%%%%%%%%%%%%%%%%%%%%%
\Omega_i(x,y;u,v)={x^i\Phi(x,u,v)-y^i\Phi(y,u,v)\over x-y}\;,
\nonumber\\
%%%%%%%%%%%%%%%%%%%%%%%%%%%%%%%%%%%%%%%%%%%%%%%%%%%%%%%%%%%%%%%%%%%%%%%%%%%
&&F_1(x,y,z,u)={1\over24}\varrho_{_{2,1}}(z,u)\Big[
{\partial^4\varrho_{_{3,1}}\over\partial x^4}
+3{\partial^3\varrho_{_{2,1}}\over\partial x^3}\Big](y,x)
-\Big\{{1\over8}{\partial\varrho_{_{2,1}}\over\partial z}
-{1\over24}{\partial^2\varrho_{_{3,1}}\over\partial z^2}
\nonumber\\
&&\hspace{2.8cm}
-{3x\over32}{\partial^2\varrho_{_{2,1}}\over\partial z^2}
+{x\over16}{\partial^3\varrho_{_{3,1}}\over\partial z^3}
-{x\over128}{\partial^4\varrho_{_{4,1}}\over\partial z^4}\Big\}(z,u)
\Big\{{\partial^4\varrho_{_{4,1}}\over\partial x^4}
-3{\partial^3\varrho_{_{3,1}}\over\partial x^3}\Big\}(x,y)
\nonumber\\
&&\hspace{2.8cm}
-{1\over18}{\partial^4\varrho_{_{4,1}}\over\partial x^4}(x,y)
+{1\over24}{\partial^3\varrho_{_{3,1}}\over\partial x^3}(x,y)
-{1\over4}{\partial^2\varrho_{_{2,1}}\over\partial x^2}(x,y)
\nonumber\\
&&\hspace{2.8cm}
-{1\over48}\Big\{\Big[(z+u)+2(z\ln z+u\ln u)\Big]
{\partial^4\varrho_{_{3,1}}\over\partial x^4}(x,y)
\nonumber\\
&&\hspace{2.8cm}
-2(z-u)^2(1+\varrho_{_{1,1}}(z,u)){\partial^4\varrho_{_{2,1}}
\over\partial x^4}(x,y)
\nonumber\\
&&\hspace{2.8cm}
-\Big[9(z+u)+6z\ln z-6u\ln u)\Big]
{\partial^3\varrho_{_{2,1}}\over\partial x^3}(x,y)
\nonumber\\
&&\hspace{2.8cm}
-\Big(4+18Q_{_\beta}+6(2-Q_{_\beta})\ln u\Big)
{\partial^2\varrho_{_{2,1}}\over\partial x^2}(x,y)
\nonumber\\
&&\hspace{2.8cm}
-\Big[2+6(1-Q_{_\beta})\ln u\Big]
{\partial\varrho_{_{1,1}}\over\partial x}(x,y)
\nonumber\\
&&\hspace{2.8cm}
-6(z-u)^2(1+\varrho_{_{1,1}}(z,u))
{\partial^3\varrho_{_{1,1}}\over\partial x^3}(x,y)
\nonumber\\
&&\hspace{2.8cm}
+6\Big[(-4+2Q_{_\beta})(z+z\ln z)+(-2-2Q_{_\beta})u
\nonumber\\
&&\hspace{2.8cm}
+(1-2Q_{_\beta})u\ln u\Big]
{\partial^2\varrho_{_{1,1}}\over\partial x^2}(x,y)
\nonumber\\
%---------------------------------------------------------------------
&&\hspace{2.8cm}
+{\partial^4\over\partial x^4}\Big[(z-u)^2\Omega_{_1}
-\Omega_{_3}\Big](x,y;z,u)
\nonumber\\
%---------------------------------------------------------------------
&&\hspace{2.8cm}
-6{\partial^4\over\partial x^3\partial u}\Big[u(z-u)
\Omega_{_1}+u\Omega_{_2}\Big](x,y;z,u)
\nonumber\\
%---------------------------------------------------------------------
&&\hspace{2.8cm}
+6{\partial^4\over\partial x^2\partial u^2}\Big[u(z+u)
\Omega_{_1}-u\Omega_{_2}\Big](x,y;z,u)
\nonumber\\
%---------------------------------------------------------------------
&&\hspace{2.8cm}
-2{\partial^4\over\partial x\partial u^3}\Big[u^2(z-u)
\Omega_{_0}+u^2\Omega_{_1}\Big](x,y;z,u)
\nonumber\\
%---------------------------------------------------------------------
&&\hspace{2.8cm}
+3{\partial^3\over\partial x^3}\Big[(z-u)^2\Omega_{_0}
+4\Big(z-u\Big)\Omega_{_1}+3\Omega_{_2}\Big](x,y;z,u)
\nonumber\\
%---------------------------------------------------------------------
&&\hspace{2.8cm}
+6{\partial^3\over\partial x\partial u^2}\Big[({5\over2}-Q_{_\beta})u(z-u)
\Omega_{_0}+{3\over2}u\Omega_{_1}\Big](x,y;z,u)
\nonumber\\
%---------------------------------------------------------------------
&&\hspace{2.8cm}
-3{\partial^3\over\partial x^2\partial u}\Big[3u(z-u)\Omega_{_0}
+\Big((6-Q_{_\beta})z
\nonumber\\
&&\hspace{2.8cm}
+(11-3Q_{_\beta})u\Big)\Omega_{_1}-(6-Q_{_\beta})\Omega_{_2}\Big](x,y;z,u)
\nonumber\\
&&\hspace{2.8cm}
-3{\partial^2\over\partial x\partial u}\Big[(7-5Q_{_\beta})(z-u)
\Omega_{_0}+(1+Q_{_\beta})\Omega_{_1}\Big](x,y;z,u)
\nonumber\\
%---------------------------------------------------------------------
&&\hspace{2.8cm}
+6{\partial^2\over\partial x^2}\Big[({7\over2}-Q_{_\beta})(z-u)
\Omega_{_0}+({9\over2}-2Q_{_\beta})\Omega_{_1}\Big](x,y;z,u)\Big\}\;,
\nonumber\\
%%%%%%%%%%%%%%%%%%%%%%%%%%%%%%%%%%%%%%%%%%%%%%%%%%%%%%%%%%%%%%%%%%%%%%%%%%%
&&F_2(x,y,z,u)=
{1\over24}\varrho_{_{2,1}}(z,u)\Big[
-{\partial^4\varrho_{_{3,1}}\over\partial x^4}
+6{\partial^2\varrho_{_{1,1}}\over\partial x^2}\Big](x,y)
+\Big\{{\partial\varrho_{_{2,1}}\over\partial z}
-{1\over3}{\partial^2\varrho_{_{3,1}}\over\partial z^2}
\nonumber\\
&&\hspace{2.8cm}
-{3x\over4}{\partial^2\varrho_{_{2,1}}\over\partial z^2}
+{x\over2}{\partial^3\varrho_{_{3,1}}\over\partial z^3}
-{x\over16}{\partial^4\varrho_{_{4,1}}\over\partial z^4}\Big\}(z,u)
\Big\{{1\over8}{\partial^4\varrho_{_{4,1}}\over\partial x^4}
-{3\over4}{\partial^3\varrho_{_{3,1}}\over\partial x^3}
\nonumber\\
&&\hspace{2.8cm}
+{3\over4}{\partial^2\varrho_{_{2,1}}\over\partial x^2}\Big\}(x,y)
+(z+u)\Big\{{1\over48}{\partial^4\varrho_{_{3,1}}\over\partial x^4}
-{1\over6}{\partial^3\varrho_{_{2,1}}\over\partial x^3}
-{1\over72}{\partial^2\varrho_{_{1,1}}\over\partial x^2}
\nonumber\\
&&\hspace{2.8cm}
+{2\over9}{\partial\varrho_{_{0,1}}\over\partial x}\Big\}(x,y)
+\Big\{{1\over18}{\partial^4\varrho_{_{4,1}}\over\partial x^4}
-{3\over8}{\partial^3\varrho_{_{3,1}}\over\partial x^3}
+{11\over18}{\partial^2\varrho_{_{2,1}}\over\partial x^2}
-{11\over36}{\partial\varrho_{_{1,1}}\over\partial x}\Big\}(x,y)
\nonumber\\
&&\hspace{2.8cm}
+(z\ln z+u\ln u)
\Big[{1\over24}{\partial^4\varrho_{_{3,1}}\over\partial x^4}
-{1\over12}{\partial^3\varrho_{_{2,1}}\over\partial x^3}
-{7\over36}{\partial^2\varrho_{_{1,1}}\over\partial x^2}
+{1\over9}{\partial\varrho_{_{0,1}}\over\partial x}\Big](x,y)
\nonumber\\
&&\hspace{2.8cm}
-(z-u)^2\Big(1+\varrho_{_{1,1}}(z,u)\Big)
\Big({1\over24}{\partial^4\varrho_{_{2,1}}\over\partial y^2\partial x^2}
+{1\over9}{\partial^3\varrho_{_{1,1}}\over\partial y^2\partial x}
+{1\over36}{\partial^3\varrho_{_{1,1}}\over\partial y\partial x^2}\Big)(x,y)
\nonumber\\
&&\hspace{2.8cm}
-{1\over48}\Big\{2{\partial^4\over\partial x\partial y^3}\Big[
(z+u)\Omega_{_2}(x,y;z,u)-\Omega_{_3}(x,y;z,u)\Big]
\nonumber\\
%---------------------------------------------------------------------
&&\hspace{2.8cm}
-{\partial^4\over\partial x^2\partial y^2}\Big[
(z-u)^2\Omega_{_1}-2(z+u)\Omega_{_2}+\Omega_{_3}\Big](x,y;z,u)
\nonumber\\
%---------------------------------------------------------------------
&&\hspace{2.8cm}
-{\partial^3\over\partial x\partial y^2}\Big[{8\over3}(z-u)^2
\Omega_{_0}+{20\over3}(z+u)\Omega_{_1}-{28\over3}\Omega_{_2}\Big](x,y;z,u)
\nonumber\\
%---------------------------------------------------------------------
&&\hspace{2.8cm}
-{2\over3}{\partial^4\over\partial x^2\partial y}\Big[(z-u)^2
\Omega_{_0}-2(z+u)\Omega_{_1}+\Omega_{_2}\Big](x,y;z,u)
\nonumber\\
%---------------------------------------------------------------------
&&\hspace{2.8cm}
+4{\partial^2\over\partial x\partial y}\Big[(z+u)\Omega_{_0}
-\Omega_{_1}\Big](x,y;z,u)\Big\}\;,
\nonumber\\
%%%%%%%%%%%%%%%%%%%%%%%%%%%%%%%%%%%%%%%%%%%%%%%%%%%%%%%%%%%%%%%%%%%%%%%%%%%
&&F_3(x,y,z,u)=-{1\over16}\Big\{2(2Q_{_\beta}-1+Q_{_\beta}\ln u)
{\partial^2\varrho_{_{2,1}}\over\partial x^2}(x,y)
\nonumber\\
%---------------------------------------------------------------------
&&\hspace{2.8cm}
+2\Big(1-2Q_{_\beta}+(1-Q_{_\beta})\ln u\Big)
{\partial\varrho_{_{1,1}}\over\partial x}(x,y)
\nonumber\\
%---------------------------------------------------------------------
&&\hspace{2.8cm}
+4\Big(z-u+z\ln z-u\ln u\Big)
{\partial^2\varrho_{_{1,1}}\over\partial x^2}(x,y)
\nonumber\\
%---------------------------------------------------------------------
&&\hspace{2.8cm}
+{\partial^3\over\partial x\partial u^2}
\Big[(1-2Q_{_\beta})u\Omega_{_1}-u(z-u)\Omega_{_0}\Big](x,y;z,u)
\nonumber\\
%---------------------------------------------------------------------
&&\hspace{2.8cm}
-{\partial^2\over\partial x\partial u}\Big[(3-5Q_{_\beta})\Omega_{_1}
-(3-Q_{_\beta})(z-u)\Omega_{_0}\Big](x,y;z,u)
\nonumber\\
%---------------------------------------------------------------------
&&\hspace{2.8cm}
-{\partial^3\over\partial x^2\partial u}\Big[Q_{_\beta}\Omega_{_2}
-(Q_{_\beta}z+(2-Q_{_\beta})u)\Omega_{_1}\Big](x,y;z,u)
\nonumber\\
%---------------------------------------------------------------------
&&\hspace{2.8cm}
-2{\partial^2\over\partial x^2}\Big[\Omega_{_1}+(z-u)\Omega_{_0}\Big]
(x,y;z,u)\Big\}\;,
\nonumber\\
%%%%%%%%%%%%%%%%%%%%%%%%%%%%%%%%%%%%%%%%%%%%%%%%%%%%%%%%%%%%%%%%%%%%%%%%%%%
&&F_4(x,y,z,u)=
-{1\over12}\varrho_{_{1,1}}(z,u)
\Big\{{\partial^4\varrho_{_{3,1}}\over\partial x^4}
+3{\partial^3\varrho_{_{2,1}}\over\partial x^3}\Big\}(x,y)
-\Big\{{1\over16}{\partial^2\varrho_{_{2,1}}\over\partial z^2}
-{1\over8}{\partial\varrho_{_{1,1}}\over\partial z}
\nonumber\\
&&\hspace{2.8cm}
+{x\over16}{\partial^2\varrho_{_{1,1}}\over\partial z^2}
-{x\over16}{\partial^3\varrho_{_{2,1}}\over\partial z^3}
+{x\over96}{\partial^4\varrho_{_{3,1}}\over\partial z^4}\Big\}(z,u)
\Big\{{\partial^4\varrho_{_{4,1}}\over\partial x^4}
-3{\partial^3\varrho_{_{3,1}}\over\partial x^3}\Big\}(x,y)
\nonumber\\
&&\hspace{2.8cm}
-\Big\{{1\over12}{\partial^4\varrho_{_{3,1}}\over\partial x^4}
+{1\over2}{\partial^3\varrho_{_{2,1}}\over\partial x^3}
+{1\over2}{\partial^2\varrho_{_{1,1}}\over\partial x^2}\Big\}(x,y)
+{1-3Q_{_\beta}\over24u}{\partial\varrho_{_{1,1}}\over\partial x}(x,y)
\nonumber\\
&&\hspace{2.8cm}
+{1-Q_{_\beta}\over8}\ln z{\partial^2\varrho_{_{1,1}}\over\partial x^2}(x,y)
-{1\over8}\ln u\Big[{\partial^3\varrho_{_{2,1}}\over\partial x^3}
+(3-Q_{_\beta}){\partial^2\varrho_{_{1,1}}\over\partial x^2}\Big](x,y)
\nonumber\\
&&\hspace{2.8cm}
-{1\over48}\Big\{{\partial^4\over\partial x\partial u^3}
\Big[u(z-u)\Omega_{_0}-u\Omega_{_1}\Big](x,y;z,u)
\nonumber\\
%---------------------------------------------------------------------
&&\hspace{2.8cm}
-3(1-Q_\beta){\partial^3\over\partial x\partial u^2}
\Big[(z-u)\Omega_{_0}-\Omega_{_1}\Big](x,y;z,u)
\nonumber\\
%---------------------------------------------------------------------
&&\hspace{2.8cm}
+3(1-Q_\beta){\partial^3\over\partial x\partial z\partial u}\Big[
(z-u)\Omega_{_0}-\Omega_{_1}\Big](x,y;z,u)
\nonumber\\
%---------------------------------------------------------------------
&&\hspace{2.8cm}
-2{\partial^4\Omega_{_2}\over\partial x^4}(x,y;z,u)
+3{\partial^4\over\partial x^3\partial u}\Big[(z-u)\Omega_{_1}
-\Omega_{_2}\Big](x,y;z,u)
\nonumber\\
%---------------------------------------------------------------------
&&\hspace{2.8cm}
-6{\partial^4\over\partial x^2\partial u^2}\Big(u\Omega_{_1}
(x,y;z,u)\Big)-6{\partial^3\Omega_{_1}\over\partial x^3}(x,y;z,u)
\nonumber\\
%---------------------------------------------------------------------
&&\hspace{2.8cm}
+3{\partial^3\over\partial x^2\partial u}\Big[(3-Q_\beta)(z-u)\Omega_{_0}
+(1-Q_\beta)\Omega_{_1}\Big](x,y;z,u)
\nonumber\\
%---------------------------------------------------------------------
&&\hspace{2.8cm}
+3(1-Q_\beta){\partial^3\over\partial x^2\partial z}\Big[(z-u)\Omega_{_0}
-\Omega_{_1}\Big](x,y;z,u)\Big\}\;,
\nonumber\\
%%%%%%%%%%%%%%%%%%%%%%%%%%%%%%%%%%%%%%%%%%%%%%%%%%%%%%%%%%%%%%%%%%%%%%%%%%%
&&F_5(x,y,z,u)={1\over12}\varrho_{_{1,1}}(z,u)
\Big\{{\partial^4\varrho_{_{3,1}}\over\partial x^4}
-6{\partial^2\varrho_{_{1,1}}\over\partial x^2}\Big\}(x,y)
+\Big\{{1\over16}{\partial^2\varrho_{_{2,1}}\over\partial z^2}
-{1\over8}{\partial\varrho_{_{1,1}}\over\partial z}
\nonumber\\
%---------------------------------------------------------------------
&&\hspace{2.8cm}
+{x\over16}{\partial^2\varrho_{_{1,1}}\over\partial z^2}
-{x\over16}{\partial^3\varrho_{_{2,1}}\over\partial z^3}
+{x\over96}{\partial^4\varrho_{_{3,1}}\over\partial z^4}\Big\}(z,u)
\Big\{{\partial^4\varrho_{_{4,1}}\over\partial x^4}
-6{\partial^3\varrho_{_{3,1}}\over\partial x^3}
\nonumber\\
%---------------------------------------------------------------------
&&\hspace{2.8cm}
+6{\partial^2\varrho_{_{2,1}}\over\partial x^2}\Big\}(x,y)
+\Big\{{1\over12}{\partial^4\varrho_{_{3,1}}\over\partial x^4}
-{1\over2}{\partial^2\varrho_{_{1,1}}\over\partial x^2}\Big\}(x,y)
\nonumber\\
%---------------------------------------------------------------------
&&\hspace{2.8cm}
+{1\over24}{\partial^4\over\partial x\partial y^3}\Omega_{_2}(x,y;z,u)
-{3\over8}{\partial^3\over\partial x\partial y^2}\Omega_{_1}(x,y;z,u)
\nonumber\\
%---------------------------------------------------------------------
&&\hspace{2.8cm}
+{1\over2}{\partial^2\over\partial x\partial y}\Omega_{_0}(x,y;z,u)\;,
\nonumber\\
%%%%%%%%%%%%%%%%%%%%%%%%%%%%%%%%%%%%%%%%%%%%%%%%%%%%%%%%%%%%%%%%%%%%%%%%%%%
&&F_6(x,y,z,u)=-{1\over16}\Big\{Q_\beta\Big[{2\over u}
{\partial\varrho_{_{1,1}}\over\partial x}(x,y)
-2{\partial^3\Omega_{_1}\over\partial x^2\partial u}(x,y;z,u)
\nonumber\\
%---------------------------------------------------------------------
&&\hspace{2.8cm}
+{\partial^3\over\partial x\partial u^2}\Big((z-u)\Omega_{_0}
-\Omega_{_1}\Big)(x,y;z,u)\Big]
\nonumber\\
%---------------------------------------------------------------------
&&\hspace{2.8cm}
-Q_\alpha\Big[2{\partial^3\Omega_{_1}\over\partial x^2\partial z}(x,y;z,u)
-{\partial^3\over\partial x\partial z\partial u}
\Big((z-u)\Omega_{_0}-\Omega_{_1}\Big)(x,y;z,u)\Big]\Big\}\;,
\nonumber\\
%%%%%%%%%%%%%%%%%%%%%%%%%%%%%%%%%%%%%%%%%%%%%%%%%%%%%%%%%%%%%%%%%%%%%%%%%%%
&&F_7(x,y,z,u)=-{1\over8}\Big\{-10{\partial\varrho_{_{1,1}}\over\partial x}
(x,y)+\ln u\Big({\partial\varrho_{_{1,1}}\over\partial x}
+{\partial^2\varrho_{_{2,1}}\over\partial x^2}\Big)(x,y)
\nonumber\\
%---------------------------------------------------------------------
&&\hspace{2.8cm}
+2(z-u)\Big(1+\varrho_{_{1,1}}(z,u)\Big)
{\partial^2\varrho_{_{1,1}}\over\partial x^2}(x,y)
-{\partial^3\over\partial x\partial u^2}\Big[(zu-u^2)
\Omega_{_0}\Big](x,y;z,u)
\nonumber\\
%---------------------------------------------------------------------
&&\hspace{2.8cm}
+{1\over2}{\partial^3\over\partial x^2\partial u}\Big[(z-3u)\Omega_{_1}
-\Omega_{_2}\Big](x,y;z,u)
\nonumber\\
%---------------------------------------------------------------------
&&\hspace{2.8cm}
-{1\over2}{\partial^2\over\partial x\partial u}\Big[\Omega_{_1}
-5(z-u)\Omega_{_0}\Big](x,y;z,u)
\nonumber\\
%---------------------------------------------------------------------
&&\hspace{2.8cm}
-{\partial^2\over\partial x^2}\Big[(z-u)\Omega_{_0}
+2\Omega_{_1}\Big](x,y;z,u)\Big\}\;,
\nonumber\\
%%%%%%%%%%%%%%%%%%%%%%%%%%%%%%%%%%%%%%%%%%%%%%%%%%%%%%%%%%%%%%%%%%%%%%%%%%%
&&F_8(x,y,z,u)=-{1\over16}\Big\{{2\over u}
{\partial\varrho_{_{1,1}}\over\partial x}(x,y)
+2(\ln z-\ln u){\partial^2\varrho_{_{1,1}}\over\partial x^2}(x,y)
\nonumber\\
%---------------------------------------------------------------------
&&\hspace{2.8cm}
-{\partial^3\over\partial x^2\partial u}
\Big[\Omega_{_1}+(z-u)\Omega_{_0}\Big](x,y;z,u)
\nonumber\\
%---------------------------------------------------------------------
&&\hspace{2.8cm}
-{\partial^3\over\partial x\partial u^2}\Big[\Omega_{_1}
-(z-u)\Omega_{_0}\Big](x,y;z,u)
\nonumber\\
%---------------------------------------------------------------------
&&\hspace{2.8cm}
+{\partial^3\over\partial x^2\partial z}\Big[\Omega_{_1}
-(z-u)\Omega_{_0}\Big](x,y;z,u)
\nonumber\\
%---------------------------------------------------------------------
&&\hspace{2.8cm}
+{\partial^3\over\partial x\partial z\partial u}\Big[\Omega_{_1}
-(z-u)\Omega_{_0}\Big](x,y;z,u)\Big\}\;,
\nonumber\\
%%%%%%%%%%%%%%%%%%%%%%%%%%%%%%%%%%%%%%%%%%%%%%%%%%%%%%%%%%%%%%%%%%%%%%%%%%%
&&F_{9}(x,y,z,u)=-{1\over16}\Big\{2(2+\ln u)\Big({\partial\varrho_{_{1,1}}
\over\partial x}-{\partial^2\varrho_{_{2,1}}\over\partial x^2}\Big)(x,y)
+2{\partial^3\over\partial x\partial u^2} \Big[u\Omega_{_1}\Big](x,y;z,u)
\nonumber\\
%---------------------------------------------------------------------
&&\hspace{2.8cm}
-{\partial^3\over\partial x^2\partial u}\Big[(z-u)\Omega_{_1}
-\Omega_{_2}\Big](x,y;z,u)
+{\partial^2\over\partial x\partial u}\Big[(z-u)\Omega_{_0}
-\Omega_{_1}\Big](x,y;z,u)\Big\}\;,
\nonumber\\
%%%%%%%%%%%%%%%%%%%%%%%%%%%%%%%%%%%%%%%%%%%%%%%%%%%%%%%%%%%%%%%%%%%%%%%%%%%
&&F_{10}(x,y,z,u)=-{1\over32}\Big\{{2\over u}{\partial\varrho_{_{1,1}}
\over\partial x}(x,y)-2\Big(2+\ln z\Big)
{\partial^2\varrho_{_{1,1}}\over\partial x^2}(x,y)
-4{\partial^3\Omega_{_1}\over\partial x^2\partial u}(x,y;z,u)
\nonumber\\
%---------------------------------------------------------------------
&&\hspace{2.8cm}
-2{\partial^3\over\partial x\partial u^2}\Big[\Omega_{_1}
-(z-u)\Omega_{_0}\Big](x,y;z,u)
\nonumber\\
%---------------------------------------------------------------------
&&\hspace{2.8cm}
-{\partial^3\over\partial x^2\partial z}\Big[5\Omega_{_1}
-(z-u)\Omega_{_0}\Big](x,y;z,u)\Big\}\;,
\nonumber\\
%%%%%%%%%%%%%%%%%%%%%%%%%%%%%%%%%%%%%%%%%%%%%%%%%%%%%%%%%%%%%%%%%%%%%%%%%%%
&&P_1(x,y,z,u,w)={1\over16}\Big\{2\Big((2-Q_\beta)\ln w+1-2Q_\beta\Big)
{\varrho_{_{1,1}}(x,z)-\varrho_{_{1,1}}(y,z)\over x-y}
\nonumber\\
&&\hspace{3.2cm}
+\Big({\partial\over\partial x}+{\partial\over\partial y}\Big)^2\Big[
3{\varrho_{_{3,1}}(x,z)-\varrho_{_{3,1}}(y,z)\over x-y}
-{\varrho_{_{3,2}}(x,z)-\varrho_{_{3,2}}(y,z)\over x-y}
\nonumber\\
&&\hspace{3.2cm}
+2(u-w+u\ln u-w\ln w){\varrho_{_{2,1}}(x,z)-\varrho_{_{2,1}}(y,z)\over x-y}\Big]
\nonumber\\
&&\hspace{3.2cm}
+2\Big({\partial\over\partial x}+{\partial\over\partial y}\Big)\Big[
2(u-w+u\ln u-w\ln w){\varrho_{_{1,1}}(x,z)-\varrho_{_{1,1}}(y,z)\over x-y}
\nonumber\\
&&\hspace{3.2cm}
+2\ln w{\varrho_{_{2,1}}(x,z)-\varrho_{_{2,1}}(y,z)\over x-y}
+{\varrho_{_{2,2}}(x,z)-\varrho_{_{2,2}}(y,z)\over x-y}\Big]
\nonumber\\
&&\hspace{3.2cm}
-{\partial^2\over\partial w^2}\Big[w(u-w){\Omega_{_0}(x,z;u,w)
-\Omega_{_0}(y,z;u,w)\over x-y}
\nonumber\\
&&\hspace{3.2cm}
+w{\Omega_{_1}(x,z;u,w)-\Omega_{_1}(y,z;u,w)\over x-y}\Big]
\nonumber\\
&&\hspace{3.2cm}
+(2-Q_\beta){\partial\over\partial w}\Big[(u-w){\Omega_{_0}(x,z;u,w)
-\Omega_{_0}(y,z;u,w)\over x-y}
\nonumber\\
&&\hspace{3.2cm}
-{\Omega_{_1}(x,z;u,w)-\Omega_{_1}(y,z;u,w)\over x-y}\Big]
\nonumber\\
&&\hspace{3.2cm}
-\Big({\partial\over\partial x}+{\partial\over\partial y}\Big)^2
\Big[{\Omega_{_2}(x,z;u,w)-\Omega_{_2}(y,z;u,w)\over x-y}
\nonumber\\
&&\hspace{3.2cm}
+(u-w){\Omega_{_1}(x,z;u,w)-\Omega_{_1}(y,z;u,w)\over x-y}\Big]
\nonumber\\
&&\hspace{3.2cm}
-2\Big({\partial\over\partial x}+{\partial\over\partial y}\Big)
{\partial\over\partial w}\Big[{\Omega_{_2}(x,z;u,w)-\Omega_{_2}(y,z;u,w)\over x-y}
\nonumber\\
&&\hspace{3.2cm}
+(u+w){\Omega_{_1}(x,z;u,w)-\Omega_{_1}(y,z;u,w)\over x-y}\Big]
\nonumber\\
%---------------------------------------------------------------------
&&\hspace{3.2cm}
-2\Big({\partial\over\partial x}+{\partial\over\partial y}\Big)\Big[
{\Omega_{_1}(x,z;u,w)-\Omega_{_1}(y,z;u,w)\over x-y}
\nonumber\\
&&\hspace{3.2cm}
+(u-w){\Omega_{_0}(x,z;u,w)-\Omega_{_0}(y,z;u,w)\over x-y}\Big]\Big\}\;,
\nonumber\\
%%%%%%%%%%%%%%%%%%%%%%%%%%%%%%%%%%%%%%%%%%%%%%%%%%%%%%%%%%%%%%%%%%%%%%%%%%%
&&P_2(x,y,z,u,w)={1\over16}\Big\{2\Big(\ln w-3+2Q_\beta-(1-Q_\beta)\ln u)\Big)
{\varrho_{_{1,1}}(x,z)-\varrho_{_{1,1}}(y,z)\over x-y}
\nonumber\\
&&\hspace{3.2cm}
+\Big({\partial\over\partial x}+{\partial\over\partial y}\Big)^2\Big[
-3{\varrho_{_{3,1}}(x,z)-\varrho_{_{3,1}}(y,z)\over x-y}
+{\varrho_{_{3,2}}(x,z)-\varrho_{_{3,2}}(y,z)\over x-y}
\nonumber\\
&&\hspace{3.2cm}
+2(u-w+u\ln u-w\ln w){\varrho_{_{2,1}}(x,z)-\varrho_{_{2,1}}(y,z)
\over x-y}\Big]
\nonumber\\
&&\hspace{3.2cm}
+2\Big({\partial\over\partial x}+{\partial\over\partial y}\Big)\Big(
2(u-w+u\ln u-w\ln w){\varrho_{_{1,1}}(x,z)-\varrho_{_{1,1}}(y,z)\over x-y}
\nonumber\\
&&\hspace{3.2cm}
-{\varrho_{_{2,2}}(x,z)-\varrho_{_{2,2}}(y,z)\over x-y}\Big)
\nonumber\\
&&\hspace{3.2cm}
-{\partial^2\over\partial w^2}\Big[w(u-w){\Omega_{_0}(x,z;u,w)
-\Omega_{_0}(y,z;u,w)\over x-y}
\nonumber\\
&&\hspace{3.2cm}
-w{\Omega_{_1}(x,z;u,w)-\Omega_{_1}(y,z;u,w)\over x-y}\Big]
\nonumber\\
%---------------------------------------------------------------------
&&\hspace{3.2cm}
+3{\partial\over\partial w}\Big[(u-w){\Omega_{_0}(x,z;u,w)
-\Omega_{_0}(y,z;u,w)\over x-y}
\nonumber\\
&&\hspace{3.2cm}
-{\Omega_{_1}(x,z;u,w)-\Omega_{_1}(y,z;u,w)\over x-y}\Big]
\nonumber\\
%---------------------------------------------------------------------
&&\hspace{3.2cm}
+\Big({\partial\over\partial x}+{\partial\over\partial y}\Big)^2\Big[
{\Omega_{_2}(x,z;u,w)-\Omega_{_2}(y,z;u,w)\over x-y}
\nonumber\\
&&\hspace{3.2cm}
-(u-w){\Omega_{_1}(x,z;u,w)-\Omega_{_1}(y,z;u,w)\over x-y}\Big]
\nonumber\\
%---------------------------------------------------------------------
&&\hspace{3.2cm}
+4w\Big({\partial\over\partial x}+{\partial\over\partial y}\Big)
{\partial\over\partial w}\Big[{\Omega_{_1}(x,z;u,w)-\Omega_{_1}(y,z;u,w)
\over x-y}\Big]
\nonumber\\
%---------------------------------------------------------------------
&&\hspace{3.2cm}
-2\Big({\partial\over\partial x}+{\partial\over\partial y}\Big)\Big[
{\Omega_{_1}(x,z;u,w)-\Omega_{_1}(y,z;u,w)\over x-y}
\nonumber\\
&&\hspace{3.2cm}
+(u-w){\Omega_{_0}(x,z;u,w)-\Omega_{_0}(y,z;u,w)\over x-y}\Big]
\nonumber\\
%---------------------------------------------------------------------
&&\hspace{3.2cm}
+(1-Q_\beta){\partial\over\partial u}\Big[{\Omega_{_1}(x,z;u,w)
-\Omega_{_1}(y,z;u,w)\over x-y}
\nonumber\\
&&\hspace{3.2cm}
-(u-w){\Omega_{_0}(x,z;u,w)-\Omega_{_0}(y,z;u,w)\over x-y}\Big]\Big\}\;,
\nonumber\\
%%%%%%%%%%%%%%%%%%%%%%%%%%%%%%%%%%%%%%%%%%%%%%%%%%%%%%%%%%%%%%%%%%%%%%%%%%%
&&P_3(x,y,z,u,w)={1\over16}\Big\{-2(2+\ln w)
{\varrho_{_{1,1}}(x,z)-\varrho_{_{1,1}}(y,z)\over x-y}
\nonumber\\
&&\hspace{3.2cm}
+(1-2Q_\beta){\partial\over\partial w}
\Big[{\Omega_{_1}(x,z;u,w)-\Omega_{_1}(y,z;u,w)\over x-y}\Big]
\nonumber\\
&&\hspace{3.2cm}
+\Big(1-(u-w){\partial\over\partial w}\Big)
\Big[{\Omega_{_0}(x,z;u,w)-\Omega_{_0}(y,z;u,w)\over x-y}\Big]\Big\}\;,
\nonumber\\
%%%%%%%%%%%%%%%%%%%%%%%%%%%%%%%%%%%%%%%%%%%%%%%%%%%%%%%%%%%%%%%%%%%%%%%%%%%
&&P_4(x,y,z,u,w)={1\over16}\Big\{2\Big(2Q_\beta+\ln w-(1-Q_\beta)\ln u\Big)
{\varrho_{_{1,1}}(x,z)-\varrho_{_{1,1}}(y,z)\over x-y}
\nonumber\\
&&\hspace{3.2cm}
-\Big(Q_\beta-(u-w){\partial\over\partial w}
-(1-Q_\beta)(u-w){\partial\over\partial u}
\Big){\Omega_{_0}(x,z;u,w)-\Omega_{_0}(y,z;u,w)\over x-y}
\nonumber\\
&&\hspace{3.2cm}
-\Big({\partial\over\partial w}+(1-Q_\beta){\partial\over\partial u}\Big)
{\Omega_{_1}(x,z;u,w)-\Omega_{_1}(y,z;u,w)\over x-y}\Big\}\;,
\nonumber\\
%%%%%%%%%%%%%%%%%%%%%%%%%%%%%%%%%%%%%%%%%%%%%%%%%%%%%%%%%%%%%%%%%%%%%%%%%%%
&&P_5(x,y,z,u,w)={1\over8\sqrt{2}}\Big\{-2(2+\ln w)
{\varrho_{_{1,1}}(x,z)-\varrho_{_{1,1}}(y,z)\over x-y}
\nonumber\\
&&\hspace{3.2cm}
-{\partial\over\partial w}\Big[{\Omega_{_1}(x,z;u,w)-\Omega_{_1}(y,z;u,w)
\over x-y}
\nonumber\\
&&\hspace{3.2cm}
+(u-w){\Omega_{_0}(x,z;u,w)-\Omega_{_0}(y,z;u,w)\over x-y}\Big]\Big\}\;,
\nonumber\\
%%%%%%%%%%%%%%%%%%%%%%%%%%%%%%%%%%%%%%%%%%%%%%%%%%%%%%%%%%%%%%%%%%%%%%%%%%%
&&P_6(x,y,z,u,w)=-{1\over8\sqrt{2}}\Big\{\Big\{2(\ln u-\ln w){\varrho_{_{1,1}}(x,z)
-\varrho_{_{1,1}}(y,z)\over x-y}
\nonumber\\
&&\hspace{3.2cm}
+\Big({\partial\over\partial u}+{\partial\over\partial w}\Big)\Big[
{\Omega_{_1}(x,z;u,w)-\Omega_{_1}(y,z;u,w)
\over x-y}\Big]
\nonumber\\
&&\hspace{3.2cm}
-(u-w)\Big({\partial\over\partial u}+{\partial\over\partial w}\Big)\Big[
{\Omega_{_0}(x,z;u,w)-\Omega_{_0}(y,z;u,w)\over x-y}\Big]\Big\}\;.
%%%%%%%%%%%%%%%%%%%%%%%%%%%%%%%%%%%%%%%%%%%%%%%%%%%%%%%%%%%%%%%%%%%%%%%%%%%
\label{eq15}
\end{eqnarray}

%%%%%%%%%%%%%%%%%%%%%%%%%%%%%%%%%%%%%%%%%%%%%%%%%%%%%%%%%%%%%%%%%%%
\end{document}